\documentclass[aps,prb,twocolumn,superscriptaddress]{revtex4-2}
\usepackage{graphicx}
\usepackage{graphicx} 

\usepackage{booktabs} 
\usepackage{array} 
\usepackage{paralist} 
\usepackage{verbatim} 
\usepackage{subfig} 
\usepackage{amssymb}
\usepackage{amsmath}
\usepackage[dvipsnames]{xcolor}
\usepackage{tikz}
\usepackage{soul}

\begin{document}
\title{A method for determining the Zeeman splitting of a spin qubit via Rabi-driven tunneling}
\author{Emily Townsend}
\email[]{emily.townsend@nist.gov}
\affiliation{Nanoscale Device Characterization Division and Joint Quantum Institute, National Institute of Standards and Technology, Gaithersburg, Maryland 20899-8423, USA; and University of Maryland, College Park, Maryland 20742, USA}
\author{Joshua Pomeroy}
\affiliation{Nanoscale Device Characterization Division , National Institute of Standards and Technology, Gaithersburg, Maryland 20899-8423, USA}
\author{Garnett W. Bryant}
\affiliation{Nanoscale Device Characterization Division  and Joint Quantum Institute, National Institute of Standards and Technology, Gaithersburg, Maryland 20899-8423, USA; and University of Maryland, College Park, Maryland 20742, USA}

\date{\today}

\begin{abstract}
We show that resonant driving between the spin up and spin down states of an electron spin-qubit in a quantum dot reduces the occupancy of the dot through leakage to an appropriately tuned lead.
A nearby charge sensor measuring the occupancy of the dot should be able to detect the narrow resonant condition upon sweeping the driving frequency.
 The presence of interactions with noisy and thermal environments such as leads are an inherent part of creating and manipulating quantum information processing systems. This work is an example of how to incorporate methods from the ``open quantum systems'' perspective into descriptions of quantum devices.
 The method used here captures the coherent dynamics of the Rabi driving on the spin state of an electron tunneling into a lead, and yields full equations of motion for the  interaction-picture density matrix with on- or off-resonance driving
\end{abstract}


\maketitle

\section{\label{Introduction}Introduction}

Modern quantum technologies exploit the ability of a quantum system to maintain its coherence while being controlled by an external routine or process.  They require isolation of the quantum system from its environment to the extent possible, as well as error correction or other ways of ameliorating environmental interference with the operation of the device.  A description of such a device based only on the quantum mechanics of the device is inherently lacking.  It is crucial to also describe the interactions of the quantum system with the elements of its environment, known and unknown, controlled and uncontrolled.
This is the regime of open quantum systems \cite{Breuer2002, Rivas2012, Lidar2020}.  In this spirit, here we attempt to understand the behavior of a simple, driven, open quantum system in a regime in which its behavior is not obvious.

Semiconductor spin qubits are a promising technology for building a quantum computer and studying open quantum systems, leveraging the advantages of a mature semiconductor processing industry and long coherence times. \cite{Burkard2023}  Such a qubit consists of an unpaired electron confined in a quantum dot, either gate-defined or dopant-based.  A static magnetic field splits the energies of the spin-up and spin-down states.  A known state for the spin can be prepared by loading an electron in the lower energy state through  tunnel coupling with a lead at an appropriate electrochemical potential relative to the electrochemical potential for those two states, then shifting the dot electrochemical potential such that neither the spin up nor spin down state will tunnel off. \cite{Elzerman2004} Subsequently an alternating current (AC) magnetic field can be used to manipulate the quantum state of the system and prepare an arbitrary coherent superposition of spin-up and spin-down.  This is known as electron spin resonance (ESR).  However successful manipulation is dependent on knowing the splitting between the spin-up and spin-down states.   The spin g-factor that relates the splitting to the strength of the magnetic field varies from device to device, as it is sensitive to band structure, interfaces between different materials in the device, impurities and electric fields. \cite{Burkard2023, Ruskov2018, Martinez2022, Cifuentes2024, Sharma2024}  
We thus seek a way to determine the resonant frequency for ESR spin flips. 

We examine a single spin qubit, driven by a microwave field, in tunneling contact with a macroscopic lead such that, sans driving, one spin state is occupied and the other unoccupied.  Typically, in almost every case, manipulation of the qubit spin state is done while tunneling is not possible for either spin state.  By combining electron spin-resonance (ESR) driving with selective tunneling, electrons in the lower energy state are driven to a state that can tunnel off, being replaced by a new electron in the lower energy state, and  the average occupation of the dot is reduced.

The experimental proposal we examine here has been described previously by Martin et al. \cite{Martin2003}, and is similar to proposals by Engel and Loss  \cite{Engel2001, Engel2002}, however we have substantially improved the theoretical description by including the coherent effect of the oscillating microwave field in the dissipative description of the tunneling.   In this work we examine the effects of the more accurate Lindblad equation that results.  This  Lindblad equation is	 valid when the driving is on or off resonance, and reveals an Autler-Townes-like splitting of the qubit levels.  Our equations are qualitatively different from those of Martin   et al., but both predict a reduced occupation of the dot when driven on resonance.  There has been at least one attempt to make the proposed measurement \cite{Xiao2004}, but it has been suggested \cite{Kouwenhoven2006, Koppens2006} that this measurement was confounded by other effects of the microwave field, specifically photon assisted tunneling (in which the microwave field induces a voltage oscillation in the lead, providing additional energy for tunneling to a state otherwise not accessible \cite{Kouwenhoven1994}) and thermal excitation.  So the requirements for experimental observation include, at the least, maintaining the electron temperature in the lead below the Zeeman splitting energy and preventing the voltage signal from appearing in the lead.  

A similar, alternative protocol currently in use adds a second quantum dot with a different g-factor initialized in a known spin state.  This allows an ESR (or EDSR, electric dipole spin resonance) manipulation on the spin in the first dot without disrupting the known state in the second dot.  Pauli spin blockade then restricts tunneling from the first dot to the second, allowing a measurement of the spin in the second dot. \cite{Koppens2006, Kouwenhoven2006, Pioro-Ladriere2008, Petersson2012, Hao2014, Maurand2016, Huang2019}  Continuous tunneling from the lead into the driven spin state is still often present in these methods, so we expect some aspects of this work to be relevant there as well. This latter technique has also been explored theoretically using hierarchical equations of motion. \cite{Hou2017}.
A simpler method is to load a spin-down electron,  plunge to an electrochemical potential which doesn't allow either spin to tunnel off, then manipulate it with a microwave field of varying frequency for a known pulse length (or a Ramsey variation with two pulses and a non-interacting time in between), and finally measure the spin state with  spin to charge conversion. \cite{Pla2012,Huang2019} 
This method might require 250 measurements of the spin state at each frequency, which might take ten minutes for a typical spectrum, and cause local heating of the electron reservoir. A spectrum must be taken at each value of static magnetic field to extract the g-factor. \cite{Pla2012}  
Another method for determining the energy  splitting between spin up and spin down states is through a transport method, such  as measuring the conductance through the device to directly query the levels as a function of magnetic field  \cite{Hanson2003} .  
Acknowledging the experimental challenges of photon assisted tunneling and local heating, the method described here might sometimes be a simpler  and faster way to know when an ESR field is resonant with the splitting, requiring no measurements of conductance on and off the dot, only the presence of a lead and a charge sensor, (when the AC magnetic field is stronger than the AC voltage difference induced by it in the lead). 	

The tunnel coupling to the lead allows electrons to move between the dot and a thermal reservoir, introducing non-unitary evolution and making this an open quantum system.  The presence of a time-dependent driving field complicates the standard method for arriving at a Lindblad equation, requiring us to explicitly evaluate the commutators of time-evolved creation and destruction operators. 
 We reserve the theoretical treatment of the driven open quantum system for a related paper \cite{TownPomBryTh2025}, and in this work discuss the application of the method to the experimental goal of finding the Zeeman splitting for an actual physical device.  We show that the hypothesized effect can be seen in the Lindblad equations describing the driven tunnelling system, and that the states that are tunnelled into and out of are also effectively shifted in energy by the Rabi oscillations, similar to the Autler-Townes splitting. \cite{Autler1955, Cohen-Tannoudji1977, Cohen-Tannoudji2008}  

	In the remaining portion of this section we will describe the physical device we consider and the Hamiltonian that describes it.  Then follows a section giving a brief, non-technical, overview of the method described more fully in reference \cite{TownPomBryTh2025}.
  Next a results section shows the driven-dissipative behavior of the device, and how to identify the desired transition energy by measuring the charge occupancy.  A conclusions section ends the paper.

\subsection{Device description and Hamiltonian}
\begin{figure}
 \includegraphics[width=0.45\textwidth]{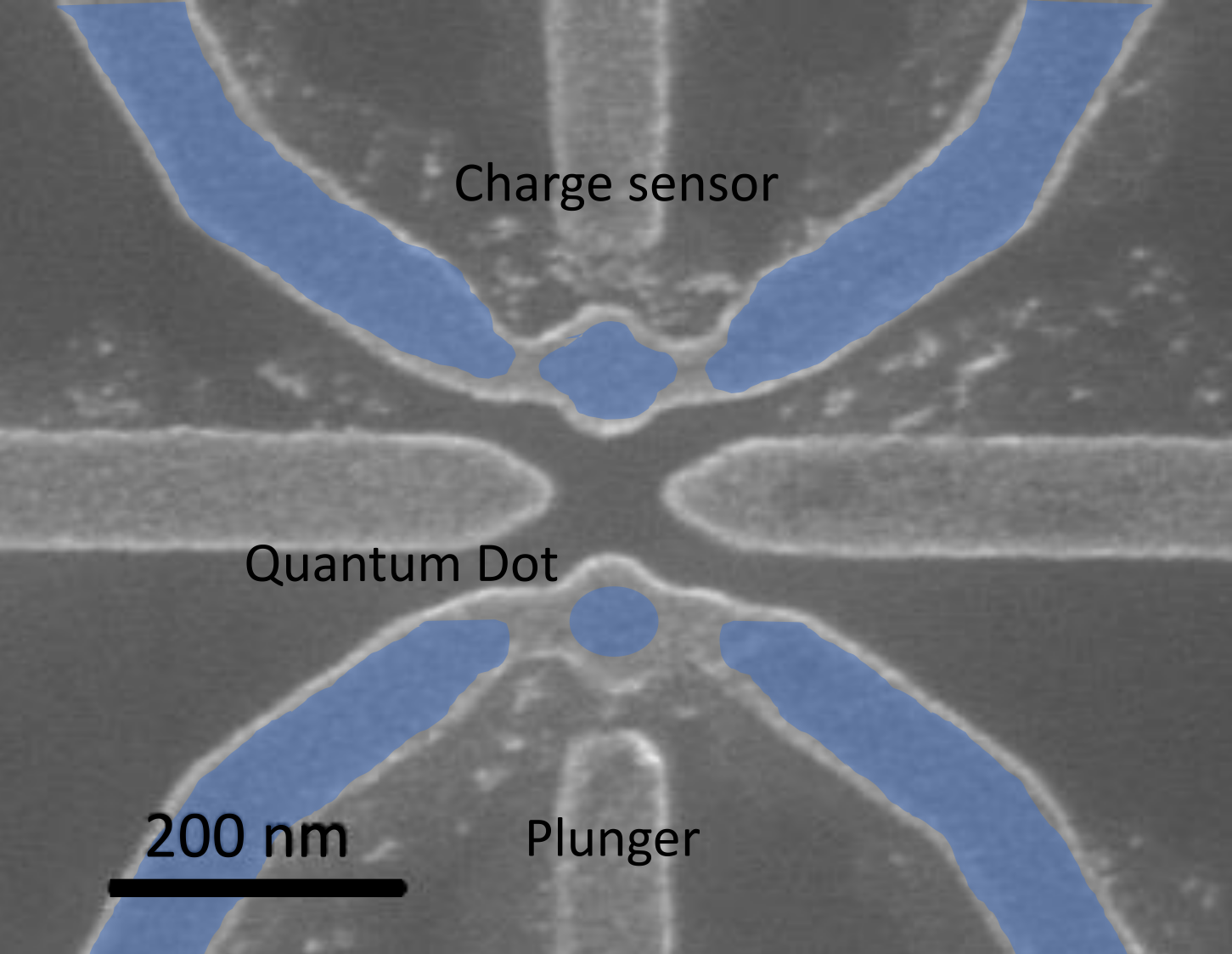}
 \caption{A scanning electron micrograph (SEM) of a device of the type considered in the paper. Dark gray regions are the surface of an oxidized silicon substrate. The lighter regions are part of a metal gate structure for generating an electron inversion layer, which is imagined as the light blue regions. The device’s lower half has the quantum dot that hosts the single spin and can exchange charge with the adjacent reservoirs via tunnelling through barriers caused by a repulsive voltage on the horizontal lead. The upper half is the charge sensor that can detect when a charge transition has occurred. }
\label{fig:GateDefDot}
\end{figure}

We consider a gate-defined dot in which an added electron can be electrostatically confined. A scanning electron micrograph of a device considered is shown in Figure \ref{fig:GateDefDot}. In this case, a pattern of metal gates (light gray) is shown on the surface of an oxidized silicon wafer (dark gray). The device has two channels (upper and lower), each of which has a continuous “U” shaped gate biased positively to create an electron gas (imagined in light blue) beneath the oxide. The upper channel is biased more positively to form a single electron transistor (SET) with sufficiently high bandwidth to sense charge transitions in the lower channel. The lower channel is used to form the quantum dot of interest, with an adjacent reservoir(s) tunnel coupled to the dot, so electrons can (un)load to(from) the dot, depending on the configuration. Gates without blue regions beneath them can be used for sculpting the electrostatic potential and manipulating the dot and SET. Finally, a microwave antenna can be fabricated above the device to provide the excitation for spin rotations.

\begin{figure}
 \includegraphics[width=0.45\textwidth]{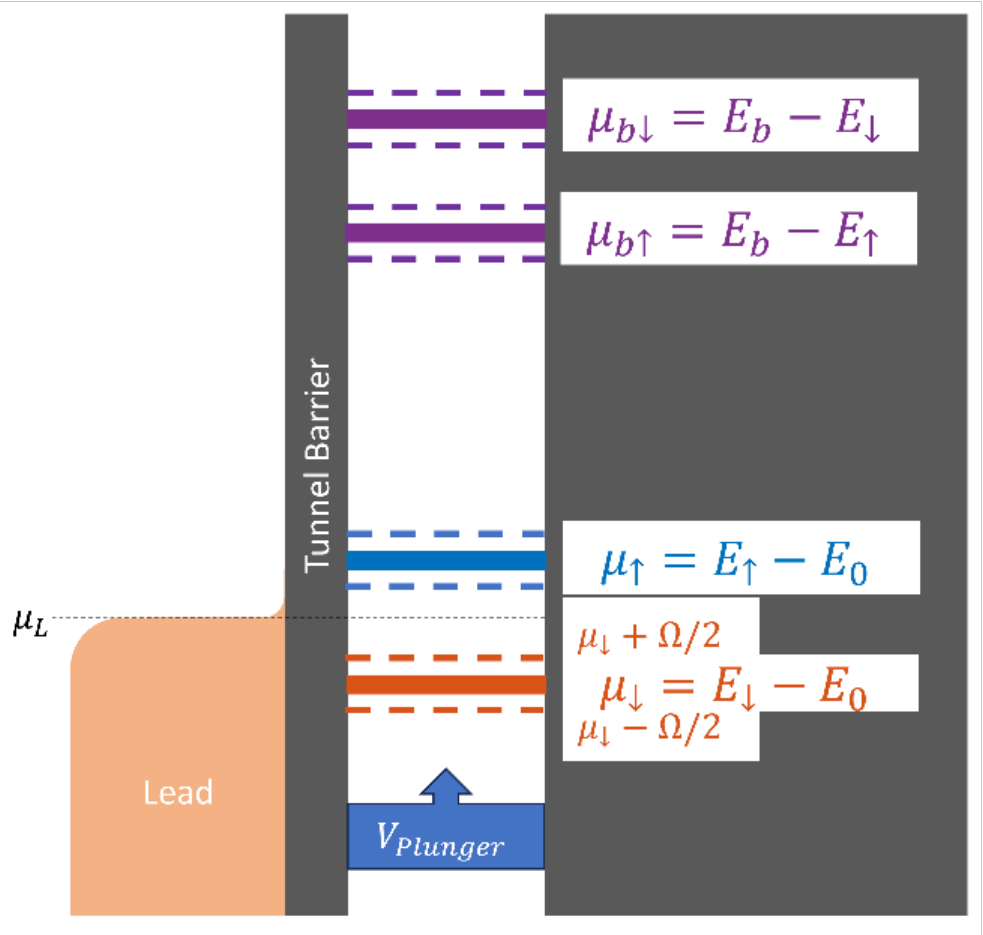}
 \caption{Relative chemical potentials of different states of the dot and the chemical potential of the lead.  There may be zero, one or two electrons on the dot.  The $1$-electron state has an unpaired spin, so spin up and spin down states are split by a static magnetic field, $E_{\uparrow} = E_1 + \hbar \omega_o/2$ and $E_{\downarrow} = E_1 - \hbar \omega_o/2$.  
	The driving, along with the Markovian approximation to the tunneling, creates the effect of transition energies that are further split by the generalized Rabi frequency: $E_{\uparrow} \pm \hbar \Omega/2$,  $E_{\downarrow} \pm \hbar \Omega/2$, $E_b - E_{\uparrow} \mp \hbar \Omega/2$,  $E_b - E_{\downarrow} \mp \hbar \Omega/2$, illustrated by dotted lines.  Results shown in this paper are all in the regime of very low temperature and a chemical potential of the lead between $ \mu_{\downarrow} +\hbar \Omega/2$  and $ \mu_{\uparrow} - \hbar \Omega/2$. }
\label{fig:muLevels}
\end{figure}

We will suppose that the first electron on the device is a single unpaired spin, which can be either in state $|\uparrow \rangle$ or $|\downarrow \rangle$, corresponding to the higher and lower energy spin states with respect to a static magnetic field.  If a second electron tunnels onto the device the state with two electrons will have no net spin magnetic moment, as both the spin up state and the spin down state will be occupied, and we refer to this state as 
 $|b\rangle$ for ``both'' spin states occupied. We include only the singlet state of  the doubly occupied dot, under the assumption that the triplet states will be at sufficiently high energy to be irrelevant. The state with no additional electron  and no net spin magnetic moment will be labeled 
 $|0\rangle$.   The differences in the energies of these states are depicted in figure  \ref{fig:muLevels}. 

We call the device holding the electrons the system, and the lead which electrons can tunnel to/from we call the environment.  The system has a Hamiltonian defined by static electric and magnetic fields and an oscillating magnetic field driving the Rabi oscillations: $H_S = H_{\rm electrostatic} + H_{\rm magnetostatic} + H_{\rm AC}(t)$. (Here we neglect the oscillating electric field, which does not affect the spin state, but potentially can provide energy for otherwise inaccesible tunneling as discussed in the introduction.) The electrostatic Hamiltonian has a term quadratic in the number of electrons, $N$, due to the Coulomb interaction between the electrons, and a term linear in $N$ due to the plunger gate and confining electrostatic potential, $\phi_{ext}$ \cite{Beenakker1991, Bush2021}.  Because we are considering only three potential charge states, we simplify the electrostatic Hamiltonian by labeling the energies for 0, 1, or 2 electrons as: $E_{0} = 0$, $E_1$
, and $ E_b = 2E_1+U $, where $U = e^2/C $ is a Coulomb interaction between the two electrons of charge $e$ on the dot which has a total capacitance $C$.  
Because of the small size of the  quantum dot, this energy $U$ is the largest  energy in the system Hamiltonian.  
The magnetostatic Hamiltonian is due to a field, $B_0$ (whose direction will define the z-direction), which will split the singly occupied energy level $E_1$ into a spin-up state,  $E_\uparrow = (E_1 + \frac{\hbar \omega_o}{2})$, and a spin down state, $E_\downarrow =(E_1 - \frac{ \hbar \omega_o}{2})$, differing in energy by $\hbar \omega_o \equiv | g\mu_B B_0 |$, where $g$ is the electron g-factor and $\mu_B$ is the Bohr magneton  and $\hbar$ the reduced Planck constant.  The driving Hamiltonian is due to an applied AC magnetic field $B_1\cos \omega t \hat x $
, which drives Rabi oscillations between the two spin states (we define the energy associated with the magnitude of this AC field: $\hbar \omega_1 \equiv g\mu_B B_1 $, and the driving frequency, $\omega$).  After making the  rotating wave approximation (RWA) to remove the rapidly oscillating terms from the AC field, this is equivalent to a rotating field  $B_1 (\cos \omega t \hat x - \sin \omega t \hat y)$ and 
 the Hamiltonian for the system is:
\begin{align}
H_S(t)= (E_1- \frac{\hbar \omega_o}{2}) \hat{n}_\downarrow +(E_1 + \frac{\hbar \omega_o}{2}) \hat{n}_\uparrow
+ U \hat{n}_\uparrow\hat{n}_\downarrow \nonumber \\
+ \frac{\hbar\omega_1}{2} \left(e^{-i\omega t} \hat a^\dagger_\uparrow \hat a_\downarrow  +  e^{i\omega t} \hat a^\dagger_\downarrow \hat a_\uparrow\right)
\label{eq:SysHam}
 \end{align}

The operators $\hat a^\dagger_s, \hat a_s$ create and destroy an electron of spin $s$ on the dot, while  $\hat{n}_s = \hat a^\dagger_s \hat a_s $ counts the number of electrons in spin state $s $ of the dot.  Because of the driving field, this Hamiltonian is time-dependent. 
 This implies an absence of energy eigenstates, because energy is not conserved.  However the time evolution of any initial state density  matrix $\rho(t_0)$ is given by $\rho(t) = \mathcal U(t,t_0) \rho(t_0)\mathcal U(t_0,t)$ in which $\mathcal U(t,t_0)$ is a time-evolution operator.  This time-evolution operator obeys the equation $ H_S(t)\, \mathcal U(t,t_0)= i\frac{\partial}{\partial t} \mathcal U(t,t_0)$ 
and can be found by a series of frame transformations detailed in appendix A 
of reference \cite{TownPomBryTh2025}.
 
The electrons in the device also have the possibility of tunneling to and from the lead, so the total Hamiltonian is a sum, $H = H_S + H_E +H_I$, which includes a Hamiltonian describing the environment, $H_E$, and a tunneling Hamiltonian for the interaction between the system and its environment:

 \begin{equation}
H_{I}=  \sum_l t ( \hat a^\dagger_{l,\uparrow} \hat a_\uparrow + \hat a ^\dagger_{l,\downarrow} \hat a_\downarrow + \hat a^\dagger_\uparrow \hat a_{l,\uparrow} +  \hat a ^\dagger_\downarrow \hat a_{l,\downarrow})
\end{equation}
in which $a^\dagger_{l,s}, a_{l,s}$ create and destroy an electron of spin $s$ in (orbital) state $l$ in the lead (with energy $\varepsilon_{l,s}$), and $t$ is the tunneling strength between a state   with spin $s$ in the dot and the state $|l,s\rangle_E$ in the lead, arising from the overlap of the orbitals.
 We assume that the tunneling strengths, $t$ in this interaction Hamiltonian are all constant, having no dependence on the lead orbital or spin state that is being tunnelled into or out of, nor the number of electrons already on the dot.  
Our derivation of a Lindblad equation will rely on a Markovian approximation that applies when the tunneling overlap is small, making $H_I$ weaker than the other two terms of the Hamiltonian.

  We don't need to specify $H_E$, other than to say that $a^\dagger_{l,s}$ creates an electron in a Fermi sea of single-particle energy levels that have some density of states which we assume to be constant near the chemical potential, $D(\varepsilon_{l,s}) = D$.  The product $\Gamma = 2\pi t^2 D/\hbar$ characterizes the tunneling rate of electrons between the dot and the lead. 
We assume that the environment is in thermodynamic equilibrium, characterized completely by its chemical potential $\mu_L$ and temperature, $T$ (or $\beta = 1/k_B T$, with $k_B$ the Boltzman  constant), and can return rapidly to that equilibrium.

The system Hamiltonian leads to the system chemical potentials depicted in Figure \ref{fig:muLevels}, 
in which the chemical potential is the change in the system energy per particle: $\mu_S(N) =\frac{\partial E}{\partial N}= E_N - E_{N-1}$.  Because there are two ways the system can add the first electron (to the up state or down), there is a chemical potential associated with each of those two possible transitions \cite{Bush2021}. Likewise for the second electron, there are two  possible starting states, up or  down. 
 Without the AC driving field the probability of a tunneling event on or off the device is determined by a Fermi function comparing the relative chemical potentials of the lead, $\mu_L$, and the chemical potential for different transitions of the system from the static portion of the Hamiltonian: 
$f(\Delta E) =  (1+e^{-\beta (\Delta E - \mu_L)})^{-1}$,
 for the following possible  transtitions: $\mu_{b\uparrow}\equiv E_b - E_\uparrow$,  $\mu_{b\downarrow}\equiv E_b - E_\downarrow$, $\mu_{\uparrow}\equiv E_\uparrow - 0$, $\mu_{\downarrow}\equiv E_\downarrow - 0$. 
A primary  result of the present  work is that including the AC driving in the calculation has an effect as if both  the spin-1/2 energy levels (and the resulting chemical potentials) were  further split by 
the generalized Rabi frequency, 
$\Omega = \sqrt{\omega_1^2 + (\omega - \omega_0)^2}$, indicated by dashed lines in Figure \ref{fig:muLevels}.  
This splitting arises due to the secular approximation or time coarse graining, which is possible because the intensity of the microwave field is strong compared to the tunneling rate into the lead.  It is similar to the Autler-Townes splitting or the AC Stark effect (in which dressed states arise due to a strong-coupling between light and atomic states compared to the rate of spontaneous photon emission),  \cite{Autler1955, Cohen-Tannoudji1977, Cohen-Tannoudji2008} and to  the vacuum Rabi splitting arising in the strong-coupling (relative to qubit and cavity decay rates) regime in circuit QED. \cite{Blais2021})  Because  only the single occupied states have a net magnetic moment, only their energies have a splitting due to Rabi driving, thus any transition energy (such as a chemical potential) between a singly occupied state and a state with zero or two electrons will be doubly split.

\section{Method Overview}
In this section we give a brief overview of the method used to derive the driven dissipative equations of motion for the spin system.  
A full disucssion of the method is in reference \cite{TownPomBryTh2025}.  A Lindblad equation is a differential equation describing the time evolution of (all the elements of the density matrix of) a dissipative system.  
There exists a standard recipe for deriving a Lindblad equation from the Hamiltonian for the system, some limited information about the Hamiltonian for the environment and the Hamiltonian for the interaction between the two. \cite{Manzano2020,Lidar2020, Rivas2012, Breuer2002} One starts with the Hamiltonian description of the density matrix of the system and the environment, and moves to the interaction picture with respect to the system and environment Hamiltonians.  Then one
 expands the expression to second order in the interaction Hamiltonian and averages over the possible states of the environment, weighted appropriately for whatever thermodynamic equilibrium state describes the environment (``tracing out the environment").  

The Born, Markov and secular (or alternatively a coarse-graining of time) approximations are the primary features of this recipe.  The Born approximation assumes that the environment does not become appreciably correlated with the system at timescales relevant to  the system dynamics.  This allows truncation at second order of the expansion due to the interaction between system and bath.   The Markovian approximation assumes that correlations in the environment decay away more rapidly than timescales relevant to the system dynamics. This leads to the ``memoryless'' feature, that only the current state of the system affects its evolution due to interactions with the environment,  not earlier states of the system.
These approximations are possible because the environment is nearly unchanging as a function of time (i.e. an electron that tunnels into the lead at an energy above the chemical potential will rapidly lose energy, keeping the lead in equilibrium).
The  secular approximation is  the disregarding of very rapid fluctuations (that average to zero) in the state of the system.  All three of these approximations allow the loss of information about the system either into the environment or into unobservable high-frequency fluctuations, which is a necessity to convert  unitary Hamiltonian evolution into dissipative (irreversible) evolution.  After one applies these approximations and traces over the environment, the rates of change of the elements of the density matrix 
depend on the current state of the system 
 and the probability that the environment is in a particular configuration. 

The time-dependent Rabi Hamiltonian of Equation \ref{eq:SysHam} presents a complication to the time averaging of the secular approximation in the standard Lindblad recipe, because the system Hamiltonian does not commute with itself at different times.  
As a  result,  the expression that is averaged over time is more complicated, and non-commuting terms with arguments at different times appear in the integrals.
The methods of references \cite{Engel2001, Engel2002, Martin2003} deal with this challenge by neglecting the Rabi Hamiltonian in the derivation of the dissipative dynamics.
 In our work we instead carefully account for the different time terms and then make the usual approximations discussed above. \cite{TownPomBryTh2025}  The result is a Markovian, completely positive trace-preserving map. Markovian means  that the evolution of the system depends only on the current state of the system and the temperature and chemical potential of the thermodynamic equilibrium of the environment. A completely positive, trace-preserving map means our approximations respect the underlying unitary evolution of the joint system.  One difference between the resulting equations and the Lindblad equation for this same system with a static Hamiltonian is that the energy levels of the system with spin appear to be split by an energy equal to the generalized Rabi frequency, $\Omega$.  For a static Hamiltonian the probability of tunneling onto the dot into a particular level involves the relative chemical potentials for the lead and the chemical potential of the dot. 
There are also terms in the static case which cancel exactly, which no longer cancel in the Rabi-driven case, resulting in a more complicated Lindblad equation. 
The full Lindblad equation in terms of the Fermi occupation functions at arbirtrary temperature and chemical potential is given in reference \cite{TownPomBryTh2025}.  To apply it to our system, we assume that the bath is effectively at zero temperature, so that its Fermi function, $n_e$, is related to the Heaviside step function, $\theta$: $n_e = f(\mu, T=0) = 1 - \theta(\mu -\mu_L)$ and $n_h \equiv 1-n_e$.  We also assume that the chemical potential of the lead is (anywhere) between the energies of the Zeeman split spin-up and spin-down states, even after the additional splitting by the generalized Rabi frequency, as depicted in Figure \ref{fig:muLevels}. (Because the tunneling rate has no state dependence in this model and we are working in the zero-temperature limit, results are unchanged when the chemical potential lies anywhere between the Rabi and Zeeman split spin-up and spin-down states.) The AC Rabi field, $B_1$ is much smaller than the static Zeeman field, $B_0$, so $\omega \sim \omega_0 >> \omega_1 \sim \Omega/2$, 
 and we assume $\omega_0$ is positive.  Thus the ranking of possible transition energies (with the Autler-Townes-like splitting) is: $E_1 - \omega/2  \pm \Omega/2 < \mu_L < E_1 + \omega/2 \pm \Omega/2 < E_b - (E_1 \pm \omega/2 \pm \Omega/2)$.  

The Fermi function of the zero temperature lead evaluated at each possible transition that the dot can make is then:
\begin{align}
n_e(\mu_\downarrow  - \Omega/2) = n_e(\mu_\downarrow  + \Omega/2) =1 \\
 n_h(\mu_\uparrow + \Omega/2) = n_h(\mu_\uparrow - \Omega/2) = 1 \\
n_h(E_b - E_{1,\rm{any}}) = 1  \\
n_h(\mu_\downarrow  - \Omega/2) = n_h(\mu_\downarrow  + \Omega/2) =0 \\
 n_e(\mu_\uparrow + \Omega/2) = n_e(\mu_\uparrow - \Omega/2) = 0\\
n_e(E_b - E_{1,\rm{any}}) = 0,  
\end{align}
in which $ E_{1,\rm{any}}$ is any of $E_1 \pm \omega/2  \pm \Omega/2$.
In this limit the elements of the interaction-picture Lindblad equation of reference \cite{TownPomBryTh2025} become:
\begin{widetext}
\begin{align}
-\frac{d\tilde{\rho}_{00}}{dt} 
&= \Gamma \left(\tilde{\rho}_{00} 
-\frac{1}{2}\left(1+\frac{a^2}{\Omega^2}\right)\tilde{\rho}_{\uparrow \uparrow}
- \frac{\omega_1^2}{2\Omega^2}\tilde{\rho}_{\downarrow \downarrow} 
-  \frac{a\omega_1}{2\Omega^2}\tilde{\rho}_{\downarrow \uparrow}
-  \frac{a\omega_1}{2\Omega^2}\tilde{\rho}_{\uparrow \downarrow} 
\right) \qquad  \mathrm{(off \, resonance)} 
\label{eq:rho00OFF} \\
&= \Gamma \left(\tilde{\rho}_{00} 
-\frac{1}{2}\tilde{\rho}_{\uparrow \uparrow} - \frac{1}{2}\tilde{\rho}_{\downarrow \downarrow} 
\right)  \qquad \mathrm{(on \, resonance)}
\\
%
-\frac{d\tilde{\rho}_{\uparrow \uparrow}}{dt}
&= 
 \Gamma \left(  \frac{1}{2}\left( 1+\frac{a^2}{\Omega^2} \right) \mathrm{\tilde{\rho}_{\uparrow \uparrow}}
 + \frac{a\omega_1}{2\Omega^2} \mathrm{\tilde{\rho}_{\uparrow \downarrow}}
+\frac{a\omega_1}{2\Omega^2}  \mathrm{\tilde{\rho}_{\downarrow \uparrow}}
-\frac{\omega_1^2}{2\Omega^2} \mathrm{\tilde{\rho}_{00}}
- \mathrm{\tilde{\rho}_{bb}} 
\right)
\qquad  \mathrm{(off \,  resonance)}\label{eq:rhoupupOFF} \\
&= 
 \Gamma \left( \mathrm{\tilde{\rho}_{\uparrow \uparrow}} \frac{1}{2}
- \mathrm{\tilde{\rho}_{00}}
\frac{1}{2}  
- \mathrm{\tilde{\rho}_{bb}}
\right) \qquad  \mathrm{(on  \, resonance)}
\\
%
-\frac{d\tilde{\rho}_{\downarrow \downarrow }}{dt} 
&= \Gamma 
\left(  \frac{\omega_1^2}{2\Omega^2} \tilde{\rho}_{\downarrow \downarrow}\
+\frac{a \omega_1}{4\Omega^2}\tilde{\rho}_{\uparrow \downarrow} 
	+\frac{a \omega_1}{4\Omega^2}\tilde{\rho}_{\downarrow \uparrow}
	- \frac{1}{2}(1+\frac{a^2}{\Omega^2})\tilde{\rho}_{00}
-\tilde{\rho}_{bb} 
\right) \qquad \mathrm{(off \, resonance)}\label{eq:rhodowndownOFF}\\
&= \Gamma 
\left(\frac{1}{2}\tilde{\rho}_{\downarrow \downarrow} 
-\frac{1}{2} \tilde{\rho}_{00} 
-\tilde{\rho}_{bb} 
\right)  \qquad \mathrm{(on \, resonance)}
\label{eq:ZeroTempDiffEqdd}
\\
%
%
-\frac{d\tilde{\rho}_{bb}}{dt} 
&= \Gamma \,  \left(
2 \tilde{\rho}_{bb}
\right) \qquad \mathrm{(on\, or\, off \,resonance)}\label{eq:rhobbOFF}
\\
%
-\frac{d \tilde{\rho}_{\uparrow \downarrow}}{dt}
&=
 \Gamma  
\left(\frac{\tilde{\rho}_{\uparrow \downarrow}}{2}
+\frac{a \omega_1}{4\Omega^2}\tilde{\rho}_{\uparrow \uparrow}
+\frac{a \omega_1}{4\Omega^2}\tilde{\rho}_{\downarrow \downarrow} 
+\frac{a \omega_1}{2\Omega^2}\tilde{\rho}_{00} 
\right)  \qquad \mathrm{(off \, resonance)} \label{eq:rhoupdownOFF}
\\
&=
 \Gamma 
\frac{\tilde{\rho}_{\uparrow \downarrow}}{2}	
 \qquad \mathrm{(on\, resonance)}.
 \label{eq:ZeroTempDiffEqUD}
\end{align}
\end{widetext}
In these equations the elements of the interaction-picture density matrix are denoted $\tilde{\rho}_{ss'} \equiv \langle s |\mathcal{U}(0,t)\rho (t) \, \mathcal{U}(t,0)|s'\rangle $, the detuning between the driving frequency and the Zeeman splitting is $a =  \omega - \omega_0$,  $\omega_0$  and $\omega_1$ are related to the static and AC magnetic fields: $\hbar \omega_{0(1)} = g\mu_B  B_{0(1)}$, and  $\Omega = \sqrt{a^2+ \omega_1^2}$ is the generalized Rabi frequency.
Taking the limit $\omega_1 \rightarrow0$, $\Omega \rightarrow a$  will recover the master equations given in reference \cite{Martin2003}.\footnote{Martin et al.  designate spin up as their lower energy state, so one must also swap up and down indices.  They also assume the double occupancy state, $b$ never occurs. }
These equations remain in the interaction picture with respect to the system Hamiltonian in equation \ref{eq:SysHam}. 
They represent the evolution due to the tunneling, but not directly due to the system Hamiltonian. 
The interaction picture Lindblad equation $\frac{d \tilde{\rho}}{dt}  = \mathcal{L}\tilde{\rho}$ transforms as:
\begin{align}
\frac{d \rho}{dt} = \mathcal{U}(t,0) \mathcal{L} \tilde{\rho}\, \mathcal{U}(0,t) - \frac{i}{\hbar} [H_S, \rho],
\end{align}
in which 
$\mathcal{U}(t,0)$ is the unitary evolution operator for the time-dependent system Hamiltonian:
\begin{align}
\frac{-i}{\hbar} H_S\, \mathcal U(t,0)&= \frac{\partial}{\partial t} \mathcal U(t,0) \label{eq:HU}
\end{align}
given in reference \cite{TownPomBryTh2025}.  

When the driving frequency is resonant with the Zeeman splitting this last term contributes 
\begin{align}
 - \frac{i}{\hbar} [H_S^{\mathrm{on-res}}, \rho] = -i\frac{\omega_1}{2} (\rho_{\uparrow \downarrow}  e^{-i\omega t} - \rho_{\downarrow \uparrow}  e^{i\omega t} )|\uparrow \rangle\langle \uparrow| \nonumber \\
  +i\frac{ \omega_1}{2} (\rho_{\uparrow \downarrow}  e^{-i\omega t} - \rho_{\downarrow \uparrow}  e^{i\omega t} )|\downarrow \rangle\langle \downarrow| \nonumber \\
    +i(\frac{ \omega_1}{2} e^{-i\omega t} (\rho_{\uparrow \uparrow}  - \rho_{\downarrow \downarrow}   )- \omega \rho_{\uparrow \downarrow})|\uparrow \rangle\langle \downarrow| \nonumber \\
      -i(\frac{ \omega_1}{2} e^{-i\omega t} (\rho_{\uparrow \uparrow}  - \rho_{\downarrow \downarrow}   )+ \omega \rho_{\downarrow \uparrow})|\downarrow \rangle\langle \uparrow|
\end{align}
By  comparing  these equations to reference \cite{Martin2003}, we can see that in that work they took the dissipative equations unaltered by the  driving, and combined them with a coherent, on-resonance driving from the system Hamiltonian.  

 An alternative to transforming the equations is to transform their solution, $\tilde{\rho}$, which may be rotated back to the Schrodinger picture using
\begin{align}
\rho_{S} = \mathcal{U}(t,0)\tilde{\rho}_{S} \,\mathcal{U}(0,t) 
\label{eq:rhotildeS}.
\end{align}

In deriving these equations we have omitted the so-called light shift terms, which behave like an additional term in the system Hamiltonian, arising from the interaction wtih the bath.  The fact that the dissipative terms do not commute with the system time evolution makes them hard to evaluate, but they are also small compared to the rest of the system Hamiltonian, being second order in the tunneling. The oscillatory part is not limited to shifts in the Schrodinger picture, because of the nature of the time evolution operator used to enter the interaction picture.  However describing the dissipative part is sufficient for predicting the charge state of the device and the effect we describe in this  work, and from here on we drop these additional oscillatory parts. 

\section{Results}
The Lindblad equation gives us access to the time evolution of the state of the dot as well as the steady-state solutions of the driven-dissipative system.  
In this section we show that the steady-state charge occupancy of the dot is reduced by one-third of an electron when the frequency of the AC driving field is resonant with the splitting due to the static magnetic field. We start with a numerical examination of  the dynamics to give a feel for the behavior of this Lindblad equation. 
 The analytical calculation of the steady state follows.

 \subsection{Dynamics}
The full dynamics may be obtained by solving the interaction-picture differential equations (\ref{eq:rho00OFF}  - \ref{eq:ZeroTempDiffEqUD}) numerically, and then transforming back from the interaction picture.  The transformation is necessary if one wants to know the (rotating) spin state of the unpaired electron on the dot.  However the relative probability of each of the charge states of the dot and the purity of the spin state 
are unaffected by the transformation.  Here we show some examples of such solutions for physically realistic parameters.   The static magnetic  field  can  be orders of magnitude larger than the magnitude of the AC field.  For the numerics we use  a static field of $B_0 = 0.50 $ T, and a driving field of $B_1 = 4.00\times 10^{-5}$ T.  For these field values and driven  on-resonance, a complete spin flip due to the Rabi field can occur with probability one in a time  
 $ \tau_\pi =  \pi/\omega_1  = 446$ ns, or for a full rotation, $\tau_{\rm Rabi} = 2\tau_\pi = 0.892 \,\mu$s.    The premise of  this  method is that the Rabi oscillations will happen rapidly, before the tunneling occurs, and the assumption that $\tau_{\rm Rabi} << 1/\Gamma$ is used in the secular approximation that yields our Lindblad equation.  This drives our choice of $1/\Gamma = 20 \mu$s, or  $\Gamma = .05(\mu s)^{-1}$ for the tunneling rate in the numerical scenarios included here.

\begin{figure}
    \includegraphics[width=0.4\textwidth]{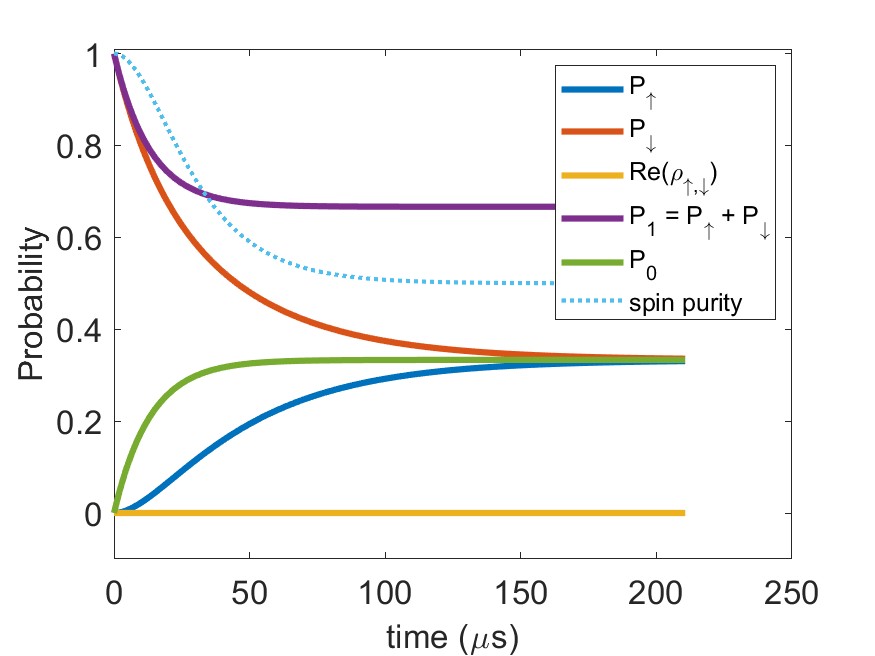}
  \includegraphics[width=0.4\textwidth]{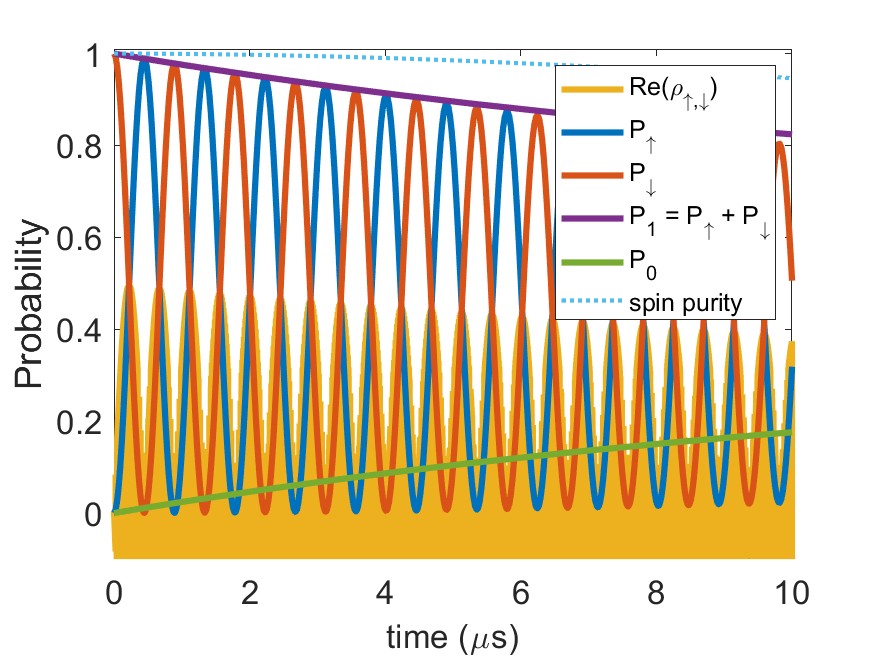}
   \includegraphics[width=0.4\textwidth]{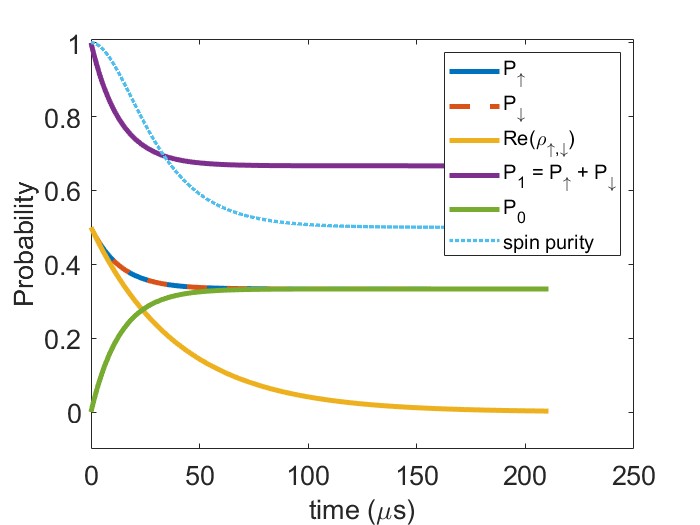}
\caption{Resonant time evolution, initial state of one electron in (a-b) the spin down direction, (c) the plus-$x$ direction. Static field $B_0 = 0.5 $ T,  AC driving field  of amplitude $B_1 = 4\times 10^{-5}$ T resonant with $B_0$ splitting, 
  tunneling to lead of  $\Gamma = .05(\mu s)^{-1}$
 .  Shown are the probabilities of finding one electron on the dot with spin up or down,  the real part of $\rho_{\uparrow \downarrow}$, the probabilities of one, two and  zero electrons, 
  and the purity of the spin state.  Figures a and c are in the interaction picture, while figure b is transformed to the lab frame, and shown for a shorter  time to resolve the oscillations.}
\label{fig:ResonantDynamics}
\end{figure}

When the generalized Rabi frequency is resonant with the splitting between the spin-up state and the spin-down state ($\omega=\omega_0$, $a=0$) the system will reach a steady state (in the interaction picture) in which it is equally likely that the dot contains a spin-up electron, contains a spin-down electron, or that the electron has tunnelled off. 
   This gives an average occupation of the dot of 2/3. 
Figure \ref{fig:ResonantDynamics} shows the dynamics of the system approaching this steady state for two initial conditions, that of a single unpaired spin-down electron on the dot (a: in the interaction picture, b: transformed into the lab frame, c: in the lab frame for a shorter time to resolve the oscillations) and an unpaired electron in the spin +$x$ state (d: in the interaction picture).  
The purity of the spin state normalized by the probability that there is an unpaired electron on the dot ($ (\rho_{\uparrow \uparrow}^2 + \rho_{\downarrow \downarrow}^2 + 2\rho_{\downarrow \uparrow} \rho_{\uparrow \downarrow})/P_N^2$) drops monotonically from its maximum value of one in the initial pure state to its minimal value of one-half at the steady-state, which is a mixed state of equal likelihood spin up and spin down. 

\clearpage

Without the driving, the steady-state of the system would be spin down, which has a purity of one.  So how does coherent driving lead to a loss of purity?  It is the combination of coherent driving with tunneling over a slower timescale the causes the loss of purity.  The tunneling is a stochastic event, that samples the state of the spin randomly as it rotates rapidly compared to the tunneling, that leads to a mixed state. 
But the tunneling into the lead has been modeled as a coherent quantum tunneling, so how does that lead to loss of purity?    Ultimately it comes from the secular approximation (which states that forcing terms which oscillate rapidly average to zero) along with the assumption that the lead (environment) is and remains in thermodynamic equilibrium.  By erasing the memory of the environment and averaging over the forcing terms we are capturing features present in a real system: We often cannot resolve small rapid changes, and even coherent evolution of a large system looks stochastic when we sample only a bit of it (or alternatively, that stochastic tunneling erases phase information; which perspective to take is a question for quantum foundations).  This is one way that a universe built on unitary quantum evolution comes to look thermal and incoherent.

\begin{figure}
  \includegraphics[width=0.45\textwidth]{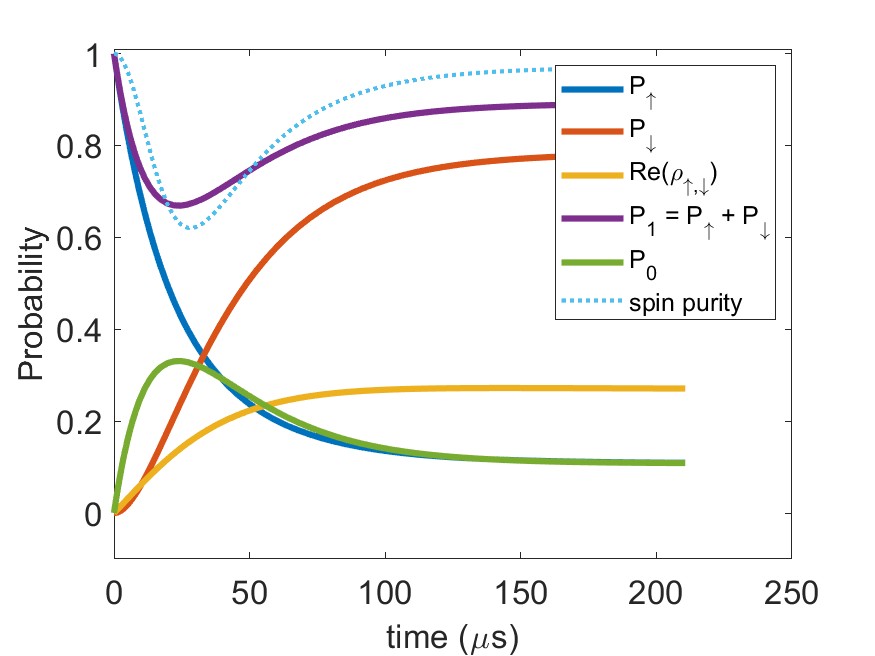}
\includegraphics[width=0.45\textwidth]{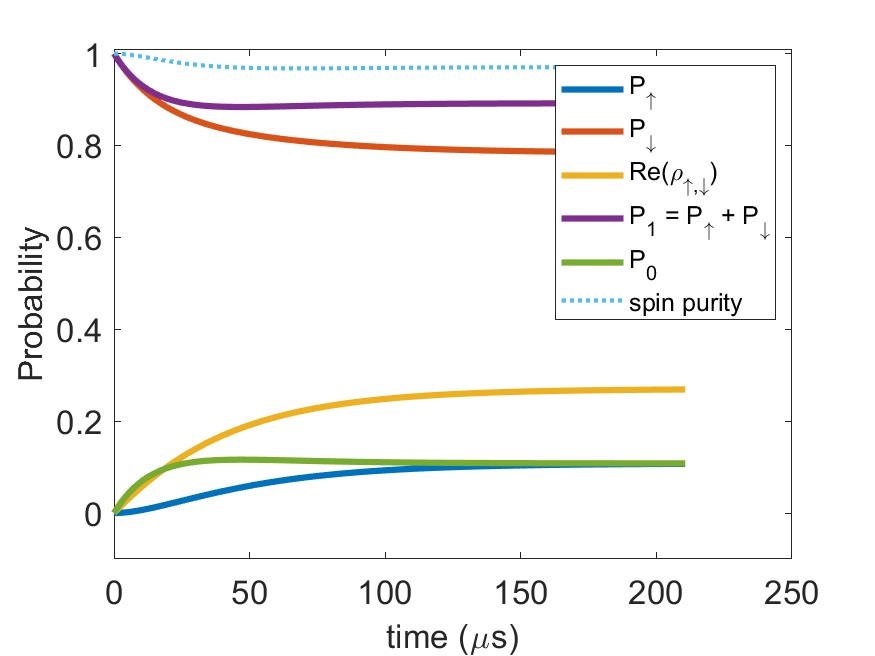}
 \caption{Time evolution in interaction picture from an initial condition of a) spin up and b) spin down, with a static field $B_0 = 0.5 $ T, and a driving field $B_1 = 4\times 10^{-5}$ T at a frequency which is $0.01\%$ below-resonance with the Zeeman splitting due to $B_0$, and a tunneling rate off the dot of $\Gamma = .05(\mu s)^{-1}$.}
\label{fig:OffRes2pDyn}
\end{figure}

Away from resonance, the steady-state average occupation of the dot is larger than the resonantly driven dot. Undriven, the dot would have an average occupation of one (as the spin-down state would be occupied). That occupation is reduced by the driving, with the maximum reduction occuring on resonance.  
 Figure \ref{fig:OffRes2pDyn} shows the dynamics of the system when the driving frequency is $0.01 \%$ below-resonant from the Zeeman splitting, and the same fields and hopping rates as the previous plot. 
 For this detuning, the time for a full Rabi rotation (two flips) is $\tau_{\rm Rabi} = 0.557 \mu s $ and the amplitude of the oscillation is $39\%$.
  Figure \ref{fig:OffRes2pDyn}  (a) shows an initial  condition of spin up, and (b) spin down.
The probability of a spin up electron being on the dot decays, while the probability that a spin down electron is on the dot and the probability that the electron has tunneled off the dot both grow.  The steady-state is an incoherent mixture of ($\sim 10\%$) having tunnelled off and ($\sim 90\%$) of the electron still on the dot in nearly a pure spin state, with a large component in the z-direction, but also some in the x-direction, as evidenced by the coherence, which is the real part of $\rho_{\uparrow \downarrow}$. (In the interaction picture there are no terms which change the phase of $\rho_{\uparrow \downarrow}$, so it remains entirely real as in the initial condition.)  The normalized purity of the spin state of the electron on the dot 
drops rapidly as the tunneling occurs, but grows again as it reaches the steady state. In the lab frame there will be rapid oscillations of the spin, rather than the steady state of the interaction picture, but again, this will have no impact on the probability of the different charge states or the normalized purity of the spin state.

Further away from resonance the Rabi oscillations are even less effective, and the steady state is primarily the spin down state.

\begin{figure}
\includegraphics[width=0.45\textwidth]{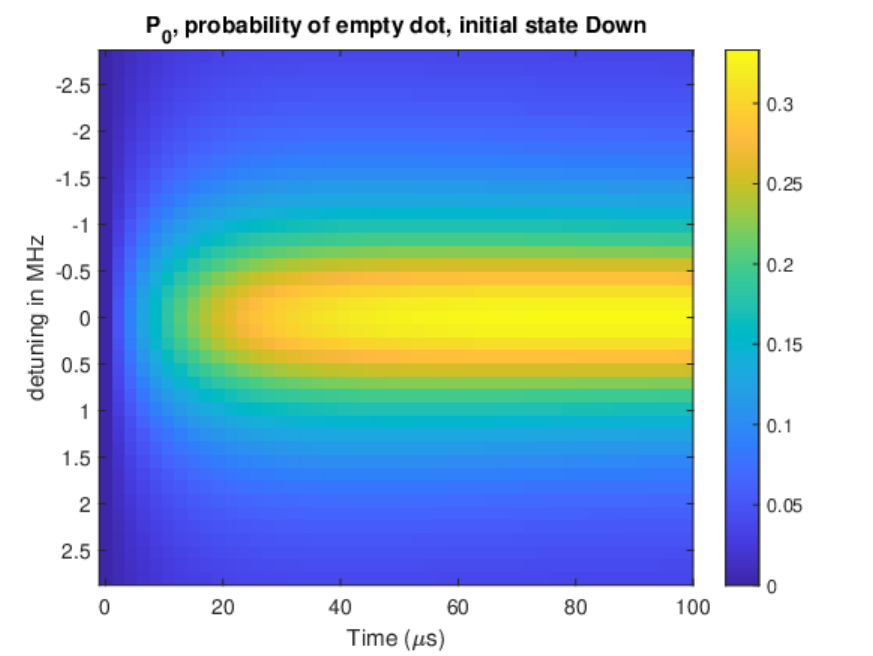}
\caption{Probability of reducted occupancy, $P_0$ as a function of time and detuning, $a= \omega -\omega_0$ for  a static field $B_0 = 0.5 $ T, and a driving field $B_1 = 4\times 10^{-5}$ T, and a tunneling rate off the dot of $\Gamma = .05(\mu s)^{-1}$, initial state $|\downarrow\rangle$.}
\label{fig:P0ofTimeDetun}
\end{figure}

To generalize the dynamics 
 to a wider range of detunings, Figure \ref{fig:P0ofTimeDetun} plots only the probability that the electron has tunnelled off, $P_0$, as a function of time and amount of detuning, for the same magnetic field strengths and tunneling rates, i.e. a static field $B_0 = 0.5 $ T, and a driving field $B_1 = 4\times 10^{-5}$ T, and a tunneling rate off the dot of $\Gamma = .05(\mu s)^{-1}$ for an initial state of one electron with spin down on the dot. 

 \subsection{Steady-state behavior}
The steady-state (in the interaction-picture)
can be obtained algebraically by setting the time derivatives to zero, which yields (for any values of $B_0, B_1, \omega$, when $\tau_{\rm Rabi}<<1/\Gamma$	):
\begin{align}
\tilde{\rho}_{\uparrow \uparrow}^{ss}  = \frac{ \omega_1^2 }{(4a^2 + 3\omega_1^2)} \\
\tilde{\rho}_{\downarrow \downarrow}^{ss}   = \frac{ (4a^2 + \omega_1^2)}{(4a^2 + 3\omega_1^2)} \nonumber \\
\tilde{\rho}_{0 0}^{ss}   =  \frac{ \omega_1^2 }{(4a^2 + 3\omega_1^2)} \\
\tilde{\rho}_{\uparrow \downarrow}^{ss} = \tilde{\rho}_{\downarrow \uparrow}^{ss} = \frac{ -2a\omega_1} {(4a^2 + 3\omega_1^2)} \nonumber \\
\tilde{\rho}_{b b}^{ss} =0  \nonumber,
\end{align}
where $a= \omega - \omega_0$ is the driving frequency detuning and $\hbar \omega_1 = g\mu_B B_1$ is determined by the magnitude of the AC driving.
This leads to the total steady-state probabilities for having zero, one, or two electrons of:
\begin{align}
P_{0} = \tilde{\rho}_{0 0}^{ss}   = \frac{ \omega_1^2 }{(4a^2 + 3\omega_1^2)}\\
P_1 = \tilde{\rho}_{\uparrow \uparrow}^{ss} + \tilde{\rho}_{\downarrow \downarrow}^{ss} = \frac{ (4a^2 + 2\omega_1^2)}{(4a^2 + 3\omega_1^2)}  \\
P_{2} = \tilde{\rho}_{b b}^{ss} =0.
\end{align}
(Moving back from the interaction picture will have no impact on the probabilities of being in the states with no spin, $P_2, P_0$, 
 nor the sum of spin up and spin down electrons, $P_1$, only the individual spin probabilities, $\rho_{\uparrow \uparrow}$ and $\rho_{\downarrow \downarrow}$.)

 This leads to a steady state expectation value of the occupancy of 
\begin{align}
\langle N_e \rangle = 1 - \frac{\omega_1^2}{4a^2 + 3\omega_1^2}
\end{align}

\begin{figure}
 \includegraphics[width=0.45\textwidth]{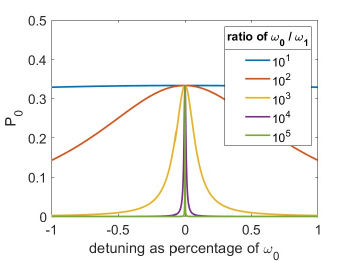}
 \includegraphics[width=0.45\textwidth]{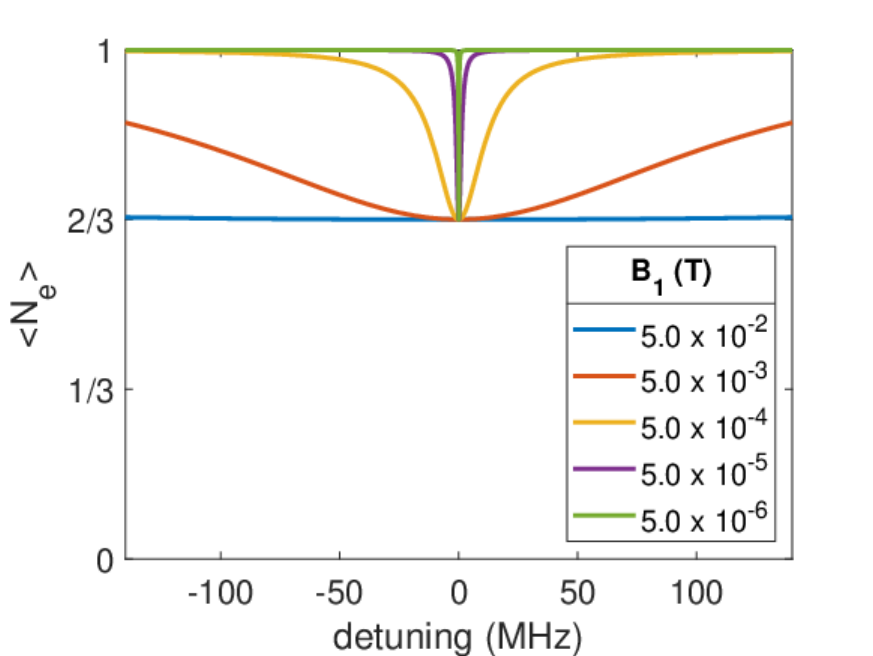}
 \caption{a) Probability that the steady state of the dot has no electrons (the unpaired spin tunnelled off) as a function of detuning percentage for various ratios of the magnitudes of the static magnetic field to the driving field.  Typical experimental ratio of $\frac{\omega_0}{\omega1}\sim 10^4$ gives a  sharp resonance. 
b) Charge occupancy of the dot as a function of absolute detuning for g = 2.002 for varying AC field strengths.}
\label{fig:WidthRes}
\end{figure}

So the steady state occupancy of the dot is reduced by one-third of an electron charge when the dot is driven on-resonance, suggesting that this could be a method for determining the resonant frequency.  The width of the resonance as a function of detuning (full width at half maximum, FWHM) is solely a function of  the  frequency derived from  the magnitude of the AC driving field, $\rm{FWHM} = \omega_1 \sqrt{3+\sqrt{21}}$.  
Figure \ref{fig:WidthRes} shows the probability of reduced occupancy in the steady state (probability of zero electrons on the dot) 
as a function of the percent detuning ($\frac{\omega - \omega_0}{\omega_0}\times 100$) for several different values of the ratio of the static and oscillating fields, as well as the average charge occupancy for the case when $B_0 =0.5T$.  The FWHM is $\sim 3 \times 10^{-4} \omega_0$ for the physically realistic case when the Zeeman field is on the order of Tesla, while the microwave field is on the order of Gauss, and the ratio is $\sim 10^4$.  

\subsection{Applicable Regime and Extensions}
These equations are applicable when the tunneling on and off the dot is much slower than the dynamics on the dot.
This also ensures that a spin-down electron has time to rotate to the up state, giving it an enhanced probability to tunnel off, before the environment of the lead measures which state it is in. The dynamics on the dot has several timescales.  The Larmor precession about the static field, $B_0$, is the fastest, with a full precession occuring in a time: $\tau_0 = 2\pi/\omega_0 = h/ g\mu_B B_0  \sim 10^{-10}  $s when $B_0 \sim 0.5$ T.  The timescale of the Rabi oscillations is set by the generalized Rabi frequency, $\Omega =  \sqrt{(\omega-\omega_0)^2 + \omega_1^2 }$.  It is slowest on resonance
,  when it is determined by the magnitude of the AC field, $B_1$.  In this case a complete pi pulse may be achieved twice  in $\tau_1 = \pi / (\omega_1) \sim h/ g\mu_B B_1 \sim  10^{-6} $s when  $B_1 = 4\times 10^{-5}$ T. If the timescale for an electron to tunnel between the dot and the lead is such that many Rabi oscillations occur, the predicted reduction of $1/3$ of an electron charge should be readily observable on resonance.   


The Lindblad approximation is a completely positive, trace-preserving map that assumes a completely memoryless reservoir, i.e. that the lead rapidly returns to equilibrium (electron filling in energy states governed by the Fermi distribution).  This would be true for macroscopic leads, however its possible that leads in atom-scale devices will retain some memory (unitary evolution of the state of the electron that tunnels into the lead).  To accurately describe that case, a more sophisticated model of the lead would be needed \cite{Rams2020, Elenewski2017, Zedler2009, deVega2017}.  
Further, we have relied upon an expectation that the temperature is well below the Zeeman splitting of the two spin states, so the lead states are either entirely filled or entirely empty of electrons at those transition energies.  If that is not the case, the differential equations governing the dynamics would be altered. 
 The full equations for arbitrary temperature can be found in our accompanying theory paper\cite{TownPomBryTh2025}.

We omitted an interaction Hamiltonian with the charge sensor.  The charge sensor will be measuring the occupancy of  the dot, preventing  a coherent superposition of different  numbers of electrons on the dot.  However this superposition is not fundamental to the operating principle,  and the resulting statistical mixture (post-measurement) is sufficient for the  effect.  The charge sensor does not measure the spin state of a singly occupied dot, so will not affect the coherent driving of the spin on the dot.

\section{Conclusion}
We used a modified Lindblad approximation to examine the driven dissipative behavior of an unpaired electron on a quantum dot with a static Zeeman magnetic field, $B_0$, and an AC Rabi driving field, 
 while it is weakly tunnel coupled to a lead, improving upon a calculation in reference \cite{Martin2003}.  We showed that the steady state charge occupancy of the dot is reduced by one-third of an electron charge when the driving frequency, $\omega$ is resonant with the Zeeman splitting, $\omega_0$ and the chemical potential of the lead is tuned to be between the two spin energy levels, shifted by the generalized Rabi frequency. 
This implies that monitoring the charge state of the dot might be used to determine the resonant frequency. The width of the resonant dip in occupation is quite narrow at physically relevant parameter regimes for spin qubits and AC driving fields, meaning that it should accurately pick out the Zeeman splitting.  This is useful because in a semiconductor device the g-factor is not known a-priori, and all gate manipulations performed on a spin qubit will be dependent upon knowing the resonant frequency. It is important that the tunneling on and off the dot be many times slower than the Rabi oscillations for these results to be applicable.
By using an open quantum systems approach we also show that driving transitions while coupled to a reservoir yields a splitting in levels similar to Autler-Townes splitting. 

 We expect this work and extensions from it  to be relevant beyond  the  single-lead, spin-1/2 quantum dot.  For example, the jump operators derived in  reference \cite{TownPomBryTh2025} can perhaps be applied when there are multiple  leads.  Beyond spin-1/2, the method could be extended to  describe the collective spin states in a multi-electron atom, allowing for an additional tool for analyzing the shell structure, with an intent similar  to reference \cite{Leon2020}.   This work might also be relevant to the study of electron spin resonance in the scanning tunneling  microscope (ESR-STM) \cite{Ast2024, Paul2016}


\section {Author Contributions}
J.P. conceived of the method and suggested this calculation to E.T.. E.T. developed the theory and performed the analytical calculations and numerical simulations in consultation with G.B..  E.T. wrote the manuscript in consultation with J.P. and G.B.. 

\section{Acknowledgements}
This research was supported in part by the National Science Foundation under Grant No. NSF PHY-1748958 and the Office of Naval Research Grant No. N00014-23-1-2477. 

%
 \bibliography{MyLibrary}

\begin{thebibliography}{40}%
\makeatletter
\providecommand \@ifxundefined [1]{%
 \@ifx{#1\undefined}
}%
\providecommand \@ifnum [1]{%
 \ifnum #1\expandafter \@firstoftwo
 \else \expandafter \@secondoftwo
 \fi
}%
\providecommand \@ifx [1]{%
 \ifx #1\expandafter \@firstoftwo
 \else \expandafter \@secondoftwo
 \fi
}%
\providecommand \natexlab [1]{#1}%
\providecommand \enquote  [1]{``#1''}%
\providecommand \bibnamefont  [1]{#1}%
\providecommand \bibfnamefont [1]{#1}%
\providecommand \citenamefont [1]{#1}%
\providecommand \href@noop [0]{\@secondoftwo}%
\providecommand \href [0]{\begingroup \@sanitize@url \@href}%
\providecommand \@href[1]{\@@startlink{#1}\@@href}%
\providecommand \@@href[1]{\endgroup#1\@@endlink}%
\providecommand \@sanitize@url [0]{\catcode `\\12\catcode `\$12\catcode
  `\&12\catcode `\#12\catcode `\^12\catcode `\_12\catcode `\%12\relax}%
\providecommand \@@startlink[1]{}%
\providecommand \@@endlink[0]{}%
\providecommand \url  [0]{\begingroup\@sanitize@url \@url }%
\providecommand \@url [1]{\endgroup\@href {#1}{\urlprefix }}%
\providecommand \urlprefix  [0]{URL }%
\providecommand \Eprint [0]{\href }%
\providecommand \doibase [0]{https://doi.org/}%
\providecommand \selectlanguage [0]{\@gobble}%
\providecommand \bibinfo  [0]{\@secondoftwo}%
\providecommand \bibfield  [0]{\@secondoftwo}%
\providecommand \translation [1]{[#1]}%
\providecommand \BibitemOpen [0]{}%
\providecommand \bibitemStop [0]{}%
\providecommand \bibitemNoStop [0]{.\EOS\space}%
\providecommand \EOS [0]{\spacefactor3000\relax}%
\providecommand \BibitemShut  [1]{\csname bibitem#1\endcsname}%
\let\auto@bib@innerbib\@empty
\bibitem [{\citenamefont {Breuer}\ and\ \citenamefont
  {Petruccione}(2002)}]{Breuer2002}%
  \BibitemOpen
  \bibfield  {author} {\bibinfo {author} {\bibfnamefont {H.-P.}\ \bibnamefont
  {Breuer}}\ and\ \bibinfo {author} {\bibfnamefont {F.}~\bibnamefont
  {Petruccione}},\ }\href@noop {} {\emph {\bibinfo {title} {The Theory of Open
  Quantum Systems}}}\ (\bibinfo  {publisher} {Oxford University Press},\
  \bibinfo {year} {2002})\BibitemShut {NoStop}%
\bibitem [{\citenamefont {Rivas}\ and\ \citenamefont
  {Huelga}(2012)}]{Rivas2012}%
  \BibitemOpen
  \bibfield  {author} {\bibinfo {author} {\bibfnamefont {{\'A}.}~\bibnamefont
  {Rivas}}\ and\ \bibinfo {author} {\bibfnamefont {S.~F.}\ \bibnamefont
  {Huelga}},\ }\href {https://doi.org/10.1007/978-3-642-23354-8} {\emph
  {\bibinfo {title} {Open Quantum Systems: An Introduction}}},\
  {{SpringerBriefs}} in {{Physics}}\ (\bibinfo  {publisher} {Springer-Verlag},\
  \bibinfo {address} {Berlin Heidelberg},\ \bibinfo {year} {2012})\BibitemShut
  {NoStop}%
\bibitem [{\citenamefont {Lidar}(2020)}]{Lidar2020}%
  \BibitemOpen
  \bibfield  {author} {\bibinfo {author} {\bibfnamefont {D.~A.}\ \bibnamefont
  {Lidar}},\ }\href {https://doi.org/10.48550/arXiv.1902.00967} {\bibinfo
  {title} {Lecture {{Notes}} on the {{Theory}} of {{Open Quantum Systems}}}}
  (\bibinfo {year} {2020}),\ \Eprint {https://arxiv.org/abs/1902.00967}
  {arXiv:1902.00967 [quant-ph]} \BibitemShut {NoStop}%
\bibitem [{\citenamefont {Burkard}\ \emph {et~al.}(2023)\citenamefont
  {Burkard}, \citenamefont {Ladd}, \citenamefont {Pan}, \citenamefont
  {Nichol},\ and\ \citenamefont {Petta}}]{Burkard2023}%
  \BibitemOpen
  \bibfield  {author} {\bibinfo {author} {\bibfnamefont {G.}~\bibnamefont
  {Burkard}}, \bibinfo {author} {\bibfnamefont {T.~D.}\ \bibnamefont {Ladd}},
  \bibinfo {author} {\bibfnamefont {A.}~\bibnamefont {Pan}}, \bibinfo {author}
  {\bibfnamefont {J.~M.}\ \bibnamefont {Nichol}},\ and\ \bibinfo {author}
  {\bibfnamefont {J.~R.}\ \bibnamefont {Petta}},\ }\href
  {https://doi.org/10.1103/RevModPhys.95.025003} {\bibfield  {journal}
  {\bibinfo  {journal} {Reviews of Modern Physics}\ }\textbf {\bibinfo {volume}
  {95}},\ \bibinfo {pages} {025003} (\bibinfo {year} {2023})}\BibitemShut
  {NoStop}%
\bibitem [{\citenamefont {Elzerman}\ \emph {et~al.}(2004)\citenamefont
  {Elzerman}, \citenamefont {Hanson}, \citenamefont {{Willems van Beveren}},
  \citenamefont {Witkamp}, \citenamefont {Vandersypen},\ and\ \citenamefont
  {Kouwenhoven}}]{Elzerman2004}%
  \BibitemOpen
  \bibfield  {author} {\bibinfo {author} {\bibfnamefont {J.~M.}\ \bibnamefont
  {Elzerman}}, \bibinfo {author} {\bibfnamefont {R.}~\bibnamefont {Hanson}},
  \bibinfo {author} {\bibfnamefont {L.~H.}\ \bibnamefont {{Willems van
  Beveren}}}, \bibinfo {author} {\bibfnamefont {B.}~\bibnamefont {Witkamp}},
  \bibinfo {author} {\bibfnamefont {L.~M.~K.}\ \bibnamefont {Vandersypen}},\
  and\ \bibinfo {author} {\bibfnamefont {L.~P.}\ \bibnamefont {Kouwenhoven}},\
  }\href {https://doi.org/10.1038/nature02693} {\bibfield  {journal} {\bibinfo
  {journal} {Nature}\ }\textbf {\bibinfo {volume} {430}},\ \bibinfo {pages}
  {431} (\bibinfo {year} {2004})}\BibitemShut {NoStop}%
\bibitem [{\citenamefont {Ruskov}\ \emph {et~al.}(2018)\citenamefont {Ruskov},
  \citenamefont {Veldhorst}, \citenamefont {Dzurak},\ and\ \citenamefont
  {Tahan}}]{Ruskov2018}%
  \BibitemOpen
  \bibfield  {author} {\bibinfo {author} {\bibfnamefont {R.}~\bibnamefont
  {Ruskov}}, \bibinfo {author} {\bibfnamefont {M.}~\bibnamefont {Veldhorst}},
  \bibinfo {author} {\bibfnamefont {A.~S.}\ \bibnamefont {Dzurak}},\ and\
  \bibinfo {author} {\bibfnamefont {C.}~\bibnamefont {Tahan}},\ }\href
  {https://doi.org/10.1103/PhysRevB.98.245424} {\bibfield  {journal} {\bibinfo
  {journal} {Physical Review B}\ }\textbf {\bibinfo {volume} {98}},\ \bibinfo
  {pages} {245424} (\bibinfo {year} {2018})}\BibitemShut {NoStop}%
\bibitem [{\citenamefont {Martinez}\ and\ \citenamefont
  {Niquet}(2022)}]{Martinez2022}%
  \BibitemOpen
  \bibfield  {author} {\bibinfo {author} {\bibfnamefont {B.}~\bibnamefont
  {Martinez}}\ and\ \bibinfo {author} {\bibfnamefont {Y.-M.}\ \bibnamefont
  {Niquet}},\ }\href {https://doi.org/10.1103/PhysRevApplied.17.024022}
  {\bibfield  {journal} {\bibinfo  {journal} {Physical Review Applied}\
  }\textbf {\bibinfo {volume} {17}},\ \bibinfo {pages} {024022} (\bibinfo
  {year} {2022})}\BibitemShut {NoStop}%
\bibitem [{\citenamefont {Cifuentes}\ \emph {et~al.}(2024)\citenamefont
  {Cifuentes}, \citenamefont {Tanttu}, \citenamefont {Gilbert}, \citenamefont
  {Huang}, \citenamefont {Vahapoglu}, \citenamefont {Leon}, \citenamefont
  {Serrano}, \citenamefont {Otter}, \citenamefont {Dunmore}, \citenamefont
  {Mai}, \citenamefont {Schlattner}, \citenamefont {Feng}, \citenamefont
  {Itoh}, \citenamefont {Abrosimov}, \citenamefont {Pohl}, \citenamefont
  {Thewalt}, \citenamefont {Laucht}, \citenamefont {Yang}, \citenamefont
  {Escott}, \citenamefont {Lim}, \citenamefont {Hudson}, \citenamefont
  {Rahman}, \citenamefont {Dzurak},\ and\ \citenamefont
  {Saraiva}}]{Cifuentes2024}%
  \BibitemOpen
  \bibfield  {author} {\bibinfo {author} {\bibfnamefont {J.~D.}\ \bibnamefont
  {Cifuentes}}, \bibinfo {author} {\bibfnamefont {T.}~\bibnamefont {Tanttu}},
  \bibinfo {author} {\bibfnamefont {W.}~\bibnamefont {Gilbert}}, \bibinfo
  {author} {\bibfnamefont {J.~Y.}\ \bibnamefont {Huang}}, \bibinfo {author}
  {\bibfnamefont {E.}~\bibnamefont {Vahapoglu}}, \bibinfo {author}
  {\bibfnamefont {R.~C.~C.}\ \bibnamefont {Leon}}, \bibinfo {author}
  {\bibfnamefont {S.}~\bibnamefont {Serrano}}, \bibinfo {author} {\bibfnamefont
  {D.}~\bibnamefont {Otter}}, \bibinfo {author} {\bibfnamefont
  {D.}~\bibnamefont {Dunmore}}, \bibinfo {author} {\bibfnamefont {P.~Y.}\
  \bibnamefont {Mai}}, \bibinfo {author} {\bibfnamefont {F.}~\bibnamefont
  {Schlattner}}, \bibinfo {author} {\bibfnamefont {M.}~\bibnamefont {Feng}},
  \bibinfo {author} {\bibfnamefont {K.}~\bibnamefont {Itoh}}, \bibinfo {author}
  {\bibfnamefont {N.}~\bibnamefont {Abrosimov}}, \bibinfo {author}
  {\bibfnamefont {H.-J.}\ \bibnamefont {Pohl}}, \bibinfo {author}
  {\bibfnamefont {M.}~\bibnamefont {Thewalt}}, \bibinfo {author} {\bibfnamefont
  {A.}~\bibnamefont {Laucht}}, \bibinfo {author} {\bibfnamefont {C.~H.}\
  \bibnamefont {Yang}}, \bibinfo {author} {\bibfnamefont {C.~C.}\ \bibnamefont
  {Escott}}, \bibinfo {author} {\bibfnamefont {W.~H.}\ \bibnamefont {Lim}},
  \bibinfo {author} {\bibfnamefont {F.~E.}\ \bibnamefont {Hudson}}, \bibinfo
  {author} {\bibfnamefont {R.}~\bibnamefont {Rahman}}, \bibinfo {author}
  {\bibfnamefont {A.~S.}\ \bibnamefont {Dzurak}},\ and\ \bibinfo {author}
  {\bibfnamefont {A.}~\bibnamefont {Saraiva}},\ }\href
  {https://doi.org/10.1038/s41467-024-48557-x} {\bibfield  {journal} {\bibinfo
  {journal} {Nature Communications}\ }\textbf {\bibinfo {volume} {15}},\
  \bibinfo {pages} {4299} (\bibinfo {year} {2024})}\BibitemShut {NoStop}%
\bibitem [{\citenamefont {Sharma}\ and\ \citenamefont
  {DiVincenzo}(2024)}]{Sharma2024}%
  \BibitemOpen
  \bibfield  {author} {\bibinfo {author} {\bibfnamefont {M.}~\bibnamefont
  {Sharma}}\ and\ \bibinfo {author} {\bibfnamefont {D.~P.}\ \bibnamefont
  {DiVincenzo}},\ }\href {https://doi.org/10.1073/pnas.2404298121} {\bibfield
  {journal} {\bibinfo  {journal} {Proceedings of the National Academy of
  Sciences}\ }\textbf {\bibinfo {volume} {121}},\ \bibinfo {pages}
  {e2404298121} (\bibinfo {year} {2024})}\BibitemShut {NoStop}%
\bibitem [{\citenamefont {Martin}\ \emph {et~al.}(2003)\citenamefont {Martin},
  \citenamefont {Mozyrsky},\ and\ \citenamefont {Jiang}}]{Martin2003}%
  \BibitemOpen
  \bibfield  {author} {\bibinfo {author} {\bibfnamefont {I.}~\bibnamefont
  {Martin}}, \bibinfo {author} {\bibfnamefont {D.}~\bibnamefont {Mozyrsky}},\
  and\ \bibinfo {author} {\bibfnamefont {H.~W.}\ \bibnamefont {Jiang}},\ }\href
  {https://doi.org/10.1103/PhysRevLett.90.018301} {\bibfield  {journal}
  {\bibinfo  {journal} {Physical Review Letters}\ }\textbf {\bibinfo {volume}
  {90}},\ \bibinfo {pages} {018301} (\bibinfo {year} {2003})}\BibitemShut
  {NoStop}%
\bibitem [{\citenamefont {Engel}\ and\ \citenamefont {Loss}(2001)}]{Engel2001}%
  \BibitemOpen
  \bibfield  {author} {\bibinfo {author} {\bibfnamefont {H.-A.}\ \bibnamefont
  {Engel}}\ and\ \bibinfo {author} {\bibfnamefont {D.}~\bibnamefont {Loss}},\
  }\href {https://doi.org/10.1103/PhysRevLett.86.4648} {\bibfield  {journal}
  {\bibinfo  {journal} {Physical Review Letters}\ }\textbf {\bibinfo {volume}
  {86}},\ \bibinfo {pages} {4648} (\bibinfo {year} {2001})}\BibitemShut
  {NoStop}%
\bibitem [{\citenamefont {Engel}\ and\ \citenamefont {Loss}(2002)}]{Engel2002}%
  \BibitemOpen
  \bibfield  {author} {\bibinfo {author} {\bibfnamefont {H.-A.}\ \bibnamefont
  {Engel}}\ and\ \bibinfo {author} {\bibfnamefont {D.}~\bibnamefont {Loss}},\
  }\href {https://doi.org/10.1103/PhysRevB.65.195321} {\bibfield  {journal}
  {\bibinfo  {journal} {Physical Review B}\ }\textbf {\bibinfo {volume} {65}},\
  \bibinfo {pages} {195321} (\bibinfo {year} {2002})}\BibitemShut {NoStop}%
\bibitem [{\citenamefont {Xiao}\ \emph {et~al.}(2004)\citenamefont {Xiao},
  \citenamefont {Martin}, \citenamefont {Yablonovitch},\ and\ \citenamefont
  {Jiang}}]{Xiao2004}%
  \BibitemOpen
  \bibfield  {author} {\bibinfo {author} {\bibfnamefont {M.}~\bibnamefont
  {Xiao}}, \bibinfo {author} {\bibfnamefont {I.}~\bibnamefont {Martin}},
  \bibinfo {author} {\bibfnamefont {E.}~\bibnamefont {Yablonovitch}},\ and\
  \bibinfo {author} {\bibfnamefont {H.~W.}\ \bibnamefont {Jiang}},\ }\href
  {https://doi.org/10.1038/nature02727} {\bibfield  {journal} {\bibinfo
  {journal} {Nature}\ }\textbf {\bibinfo {volume} {430}},\ \bibinfo {pages}
  {435} (\bibinfo {year} {2004})}\BibitemShut {NoStop}%
\bibitem [{\citenamefont {Kouwenhoven}\ \emph {et~al.}(2006)\citenamefont
  {Kouwenhoven}, \citenamefont {Elzerman}, \citenamefont {Hanson},
  \citenamefont {{Willems van Beveren}},\ and\ \citenamefont
  {Vandersypen}}]{Kouwenhoven2006}%
  \BibitemOpen
  \bibfield  {author} {\bibinfo {author} {\bibfnamefont {L.~P.}\ \bibnamefont
  {Kouwenhoven}}, \bibinfo {author} {\bibfnamefont {J.~M.}\ \bibnamefont
  {Elzerman}}, \bibinfo {author} {\bibfnamefont {R.}~\bibnamefont {Hanson}},
  \bibinfo {author} {\bibfnamefont {L.~H.}\ \bibnamefont {{Willems van
  Beveren}}},\ and\ \bibinfo {author} {\bibfnamefont {L.~M.~K.}\ \bibnamefont
  {Vandersypen}},\ }\href {https://doi.org/10.1002/pssb.200642228} {\bibfield
  {journal} {\bibinfo  {journal} {physica status solidi (b)}\ }\textbf
  {\bibinfo {volume} {243}},\ \bibinfo {pages} {3682} (\bibinfo {year}
  {2006})}\BibitemShut {NoStop}%
\bibitem [{\citenamefont {Koppens}\ \emph {et~al.}(2006)\citenamefont
  {Koppens}, \citenamefont {Buizert}, \citenamefont {Tielrooij}, \citenamefont
  {Vink}, \citenamefont {Nowack}, \citenamefont {Meunier}, \citenamefont
  {Kouwenhoven},\ and\ \citenamefont {Vandersypen}}]{Koppens2006}%
  \BibitemOpen
  \bibfield  {author} {\bibinfo {author} {\bibfnamefont {F.~H.~L.}\
  \bibnamefont {Koppens}}, \bibinfo {author} {\bibfnamefont {C.}~\bibnamefont
  {Buizert}}, \bibinfo {author} {\bibfnamefont {K.~J.}\ \bibnamefont
  {Tielrooij}}, \bibinfo {author} {\bibfnamefont {I.~T.}\ \bibnamefont {Vink}},
  \bibinfo {author} {\bibfnamefont {K.~C.}\ \bibnamefont {Nowack}}, \bibinfo
  {author} {\bibfnamefont {T.}~\bibnamefont {Meunier}}, \bibinfo {author}
  {\bibfnamefont {L.~P.}\ \bibnamefont {Kouwenhoven}},\ and\ \bibinfo {author}
  {\bibfnamefont {L.~M.~K.}\ \bibnamefont {Vandersypen}},\ }\href
  {https://doi.org/10.1038/nature05065} {\bibfield  {journal} {\bibinfo
  {journal} {Nature}\ }\textbf {\bibinfo {volume} {442}},\ \bibinfo {pages}
  {766} (\bibinfo {year} {2006})}\BibitemShut {NoStop}%
\bibitem [{\citenamefont {Kouwenhoven}\ \emph {et~al.}(1994)\citenamefont
  {Kouwenhoven}, \citenamefont {Jauhar}, \citenamefont {McCormick},
  \citenamefont {Dixon}, \citenamefont {McEuen}, \citenamefont {Nazarov},
  \citenamefont {{van der Vaart}},\ and\ \citenamefont
  {Foxon}}]{Kouwenhoven1994}%
  \BibitemOpen
  \bibfield  {author} {\bibinfo {author} {\bibfnamefont {L.~P.}\ \bibnamefont
  {Kouwenhoven}}, \bibinfo {author} {\bibfnamefont {S.}~\bibnamefont {Jauhar}},
  \bibinfo {author} {\bibfnamefont {K.}~\bibnamefont {McCormick}}, \bibinfo
  {author} {\bibfnamefont {D.}~\bibnamefont {Dixon}}, \bibinfo {author}
  {\bibfnamefont {P.~L.}\ \bibnamefont {McEuen}}, \bibinfo {author}
  {\bibfnamefont {{\relax Yu}.~V.}\ \bibnamefont {Nazarov}}, \bibinfo {author}
  {\bibfnamefont {N.~C.}\ \bibnamefont {{van der Vaart}}},\ and\ \bibinfo
  {author} {\bibfnamefont {C.~T.}\ \bibnamefont {Foxon}},\ }\href
  {https://doi.org/10.1103/PhysRevB.50.2019} {\bibfield  {journal} {\bibinfo
  {journal} {Physical Review B}\ }\textbf {\bibinfo {volume} {50}},\ \bibinfo
  {pages} {2019} (\bibinfo {year} {1994})}\BibitemShut {NoStop}%
\bibitem [{\citenamefont {{Pioro-Ladri{\`e}re}}\ \emph
  {et~al.}(2008)\citenamefont {{Pioro-Ladri{\`e}re}}, \citenamefont {Obata},
  \citenamefont {Tokura}, \citenamefont {Shin}, \citenamefont {Kubo},
  \citenamefont {Yoshida}, \citenamefont {Taniyama},\ and\ \citenamefont
  {Tarucha}}]{Pioro-Ladriere2008}%
  \BibitemOpen
  \bibfield  {author} {\bibinfo {author} {\bibfnamefont {M.}~\bibnamefont
  {{Pioro-Ladri{\`e}re}}}, \bibinfo {author} {\bibfnamefont {T.}~\bibnamefont
  {Obata}}, \bibinfo {author} {\bibfnamefont {Y.}~\bibnamefont {Tokura}},
  \bibinfo {author} {\bibfnamefont {Y.-S.}\ \bibnamefont {Shin}}, \bibinfo
  {author} {\bibfnamefont {T.}~\bibnamefont {Kubo}}, \bibinfo {author}
  {\bibfnamefont {K.}~\bibnamefont {Yoshida}}, \bibinfo {author} {\bibfnamefont
  {T.}~\bibnamefont {Taniyama}},\ and\ \bibinfo {author} {\bibfnamefont
  {S.}~\bibnamefont {Tarucha}},\ }\href {https://doi.org/10.1038/nphys1053}
  {\bibfield  {journal} {\bibinfo  {journal} {Nature Physics}\ }\textbf
  {\bibinfo {volume} {4}},\ \bibinfo {pages} {776} (\bibinfo {year}
  {2008})}\BibitemShut {NoStop}%
\bibitem [{\citenamefont {Petersson}\ \emph {et~al.}(2012)\citenamefont
  {Petersson}, \citenamefont {McFaul}, \citenamefont {Schroer}, \citenamefont
  {Jung}, \citenamefont {Taylor}, \citenamefont {Houck},\ and\ \citenamefont
  {Petta}}]{Petersson2012}%
  \BibitemOpen
  \bibfield  {author} {\bibinfo {author} {\bibfnamefont {K.~D.}\ \bibnamefont
  {Petersson}}, \bibinfo {author} {\bibfnamefont {L.~W.}\ \bibnamefont
  {McFaul}}, \bibinfo {author} {\bibfnamefont {M.~D.}\ \bibnamefont {Schroer}},
  \bibinfo {author} {\bibfnamefont {M.}~\bibnamefont {Jung}}, \bibinfo {author}
  {\bibfnamefont {J.~M.}\ \bibnamefont {Taylor}}, \bibinfo {author}
  {\bibfnamefont {A.~A.}\ \bibnamefont {Houck}},\ and\ \bibinfo {author}
  {\bibfnamefont {J.~R.}\ \bibnamefont {Petta}},\ }\href
  {https://doi.org/10.1038/nature11559} {\bibfield  {journal} {\bibinfo
  {journal} {Nature}\ }\textbf {\bibinfo {volume} {490}},\ \bibinfo {pages}
  {380} (\bibinfo {year} {2012})}\BibitemShut {NoStop}%
\bibitem [{\citenamefont {Hao}\ \emph {et~al.}(2014)\citenamefont {Hao},
  \citenamefont {Ruskov}, \citenamefont {Xiao}, \citenamefont {Tahan},\ and\
  \citenamefont {Jiang}}]{Hao2014}%
  \BibitemOpen
  \bibfield  {author} {\bibinfo {author} {\bibfnamefont {X.}~\bibnamefont
  {Hao}}, \bibinfo {author} {\bibfnamefont {R.}~\bibnamefont {Ruskov}},
  \bibinfo {author} {\bibfnamefont {M.}~\bibnamefont {Xiao}}, \bibinfo {author}
  {\bibfnamefont {C.}~\bibnamefont {Tahan}},\ and\ \bibinfo {author}
  {\bibfnamefont {H.}~\bibnamefont {Jiang}},\ }\href
  {https://doi.org/10.1038/ncomms4860} {\bibfield  {journal} {\bibinfo
  {journal} {Nature Communications}\ }\textbf {\bibinfo {volume} {5}},\
  \bibinfo {pages} {3860} (\bibinfo {year} {2014})}\BibitemShut {NoStop}%
\bibitem [{\citenamefont {Maurand}\ \emph {et~al.}(2016)\citenamefont
  {Maurand}, \citenamefont {Jehl}, \citenamefont {{Kotekar-Patil}},
  \citenamefont {Corna}, \citenamefont {Bohuslavskyi}, \citenamefont
  {Lavi{\'e}ville}, \citenamefont {Hutin}, \citenamefont {Barraud},
  \citenamefont {Vinet}, \citenamefont {Sanquer},\ and\ \citenamefont
  {De~Franceschi}}]{Maurand2016}%
  \BibitemOpen
  \bibfield  {author} {\bibinfo {author} {\bibfnamefont {R.}~\bibnamefont
  {Maurand}}, \bibinfo {author} {\bibfnamefont {X.}~\bibnamefont {Jehl}},
  \bibinfo {author} {\bibfnamefont {D.}~\bibnamefont {{Kotekar-Patil}}},
  \bibinfo {author} {\bibfnamefont {A.}~\bibnamefont {Corna}}, \bibinfo
  {author} {\bibfnamefont {H.}~\bibnamefont {Bohuslavskyi}}, \bibinfo {author}
  {\bibfnamefont {R.}~\bibnamefont {Lavi{\'e}ville}}, \bibinfo {author}
  {\bibfnamefont {L.}~\bibnamefont {Hutin}}, \bibinfo {author} {\bibfnamefont
  {S.}~\bibnamefont {Barraud}}, \bibinfo {author} {\bibfnamefont
  {M.}~\bibnamefont {Vinet}}, \bibinfo {author} {\bibfnamefont
  {M.}~\bibnamefont {Sanquer}},\ and\ \bibinfo {author} {\bibfnamefont
  {S.}~\bibnamefont {De~Franceschi}},\ }\href
  {https://doi.org/10.1038/ncomms13575} {\bibfield  {journal} {\bibinfo
  {journal} {Nature Communications}\ }\textbf {\bibinfo {volume} {7}},\
  \bibinfo {pages} {13575} (\bibinfo {year} {2016})}\BibitemShut {NoStop}%
\bibitem [{\citenamefont {Huang}\ \emph {et~al.}(2019)\citenamefont {Huang},
  \citenamefont {Yang}, \citenamefont {Chan}, \citenamefont {Tanttu},
  \citenamefont {Hensen}, \citenamefont {Leon}, \citenamefont {Fogarty},
  \citenamefont {Hwang}, \citenamefont {Hudson}, \citenamefont {Itoh},
  \citenamefont {Morello}, \citenamefont {Laucht},\ and\ \citenamefont
  {Dzurak}}]{Huang2019}%
  \BibitemOpen
  \bibfield  {author} {\bibinfo {author} {\bibfnamefont {W.}~\bibnamefont
  {Huang}}, \bibinfo {author} {\bibfnamefont {C.~H.}\ \bibnamefont {Yang}},
  \bibinfo {author} {\bibfnamefont {K.~W.}\ \bibnamefont {Chan}}, \bibinfo
  {author} {\bibfnamefont {T.}~\bibnamefont {Tanttu}}, \bibinfo {author}
  {\bibfnamefont {B.}~\bibnamefont {Hensen}}, \bibinfo {author} {\bibfnamefont
  {R.~C.~C.}\ \bibnamefont {Leon}}, \bibinfo {author} {\bibfnamefont {M.~A.}\
  \bibnamefont {Fogarty}}, \bibinfo {author} {\bibfnamefont {J.~C.~C.}\
  \bibnamefont {Hwang}}, \bibinfo {author} {\bibfnamefont {F.~E.}\ \bibnamefont
  {Hudson}}, \bibinfo {author} {\bibfnamefont {K.~M.}\ \bibnamefont {Itoh}},
  \bibinfo {author} {\bibfnamefont {A.}~\bibnamefont {Morello}}, \bibinfo
  {author} {\bibfnamefont {A.}~\bibnamefont {Laucht}},\ and\ \bibinfo {author}
  {\bibfnamefont {A.~S.}\ \bibnamefont {Dzurak}},\ }\href
  {https://doi.org/10.1038/s41586-019-1197-0} {\bibfield  {journal} {\bibinfo
  {journal} {Nature}\ }\textbf {\bibinfo {volume} {569}},\ \bibinfo {pages}
  {532} (\bibinfo {year} {2019})}\BibitemShut {NoStop}%
\bibitem [{\citenamefont {Hou}\ \emph {et~al.}(2017)\citenamefont {Hou},
  \citenamefont {Wang}, \citenamefont {Wei},\ and\ \citenamefont
  {Yan}}]{Hou2017}%
  \BibitemOpen
  \bibfield  {author} {\bibinfo {author} {\bibfnamefont {W.}~\bibnamefont
  {Hou}}, \bibinfo {author} {\bibfnamefont {Y.}~\bibnamefont {Wang}}, \bibinfo
  {author} {\bibfnamefont {J.}~\bibnamefont {Wei}},\ and\ \bibinfo {author}
  {\bibfnamefont {Y.}~\bibnamefont {Yan}},\ }\href
  {https://doi.org/10.1063/1.4985146} {\bibfield  {journal} {\bibinfo
  {journal} {The Journal of Chemical Physics}\ }\textbf {\bibinfo {volume}
  {146}},\ \bibinfo {pages} {224304} (\bibinfo {year} {2017})}\BibitemShut
  {NoStop}%
\bibitem [{\citenamefont {Pla}\ \emph {et~al.}(2012)\citenamefont {Pla},
  \citenamefont {Tan}, \citenamefont {Dehollain}, \citenamefont {Lim},
  \citenamefont {Morton}, \citenamefont {Jamieson}, \citenamefont {Dzurak},\
  and\ \citenamefont {Morello}}]{Pla2012}%
  \BibitemOpen
  \bibfield  {author} {\bibinfo {author} {\bibfnamefont {J.~J.}\ \bibnamefont
  {Pla}}, \bibinfo {author} {\bibfnamefont {K.~Y.}\ \bibnamefont {Tan}},
  \bibinfo {author} {\bibfnamefont {J.~P.}\ \bibnamefont {Dehollain}}, \bibinfo
  {author} {\bibfnamefont {W.~H.}\ \bibnamefont {Lim}}, \bibinfo {author}
  {\bibfnamefont {J.~J.~L.}\ \bibnamefont {Morton}}, \bibinfo {author}
  {\bibfnamefont {D.~N.}\ \bibnamefont {Jamieson}}, \bibinfo {author}
  {\bibfnamefont {A.~S.}\ \bibnamefont {Dzurak}},\ and\ \bibinfo {author}
  {\bibfnamefont {A.}~\bibnamefont {Morello}},\ }\href
  {https://doi.org/10.1038/nature11449} {\bibfield  {journal} {\bibinfo
  {journal} {Nature}\ }\textbf {\bibinfo {volume} {489}},\ \bibinfo {pages}
  {541} (\bibinfo {year} {2012})}\BibitemShut {NoStop}%
\bibitem [{\citenamefont {Hanson}\ \emph {et~al.}(2003)\citenamefont {Hanson},
  \citenamefont {Witkamp}, \citenamefont {Vandersypen}, \citenamefont {{van
  Beveren}}, \citenamefont {Elzerman},\ and\ \citenamefont
  {Kouwenhoven}}]{Hanson2003}%
  \BibitemOpen
  \bibfield  {author} {\bibinfo {author} {\bibfnamefont {R.}~\bibnamefont
  {Hanson}}, \bibinfo {author} {\bibfnamefont {B.}~\bibnamefont {Witkamp}},
  \bibinfo {author} {\bibfnamefont {L.~M.~K.}\ \bibnamefont {Vandersypen}},
  \bibinfo {author} {\bibfnamefont {L.~H.~W.}\ \bibnamefont {{van Beveren}}},
  \bibinfo {author} {\bibfnamefont {J.~M.}\ \bibnamefont {Elzerman}},\ and\
  \bibinfo {author} {\bibfnamefont {L.~P.}\ \bibnamefont {Kouwenhoven}},\
  }\href {https://doi.org/10.1103/PhysRevLett.91.196802} {\bibfield  {journal}
  {\bibinfo  {journal} {Physical Review Letters}\ }\textbf {\bibinfo {volume}
  {91}},\ \bibinfo {pages} {196802} (\bibinfo {year} {2003})}\BibitemShut
  {NoStop}%
\bibitem [{\citenamefont {Townsend}\ \emph {et~al.}(2025)\citenamefont
  {Townsend}, \citenamefont {Pomeroy},\ and\ \citenamefont
  {Bryant}}]{TownPomBryTh2025}%
  \BibitemOpen
  \bibfield  {author} {\bibinfo {author} {\bibfnamefont {E.}~\bibnamefont
  {Townsend}}, \bibinfo {author} {\bibfnamefont {J.~M.}\ \bibnamefont
  {Pomeroy}},\ and\ \bibinfo {author} {\bibfnamefont {G.~W.}\ \bibnamefont
  {Bryant}},\ }\href@noop {} {\bibfield  {journal} {\bibinfo  {journal} {TBP}\
  } (\bibinfo {year} {2025})}\BibitemShut {NoStop}%
\bibitem [{\citenamefont {Autler}\ and\ \citenamefont
  {Townes}(1955)}]{Autler1955}%
  \BibitemOpen
  \bibfield  {author} {\bibinfo {author} {\bibfnamefont {S.~H.}\ \bibnamefont
  {Autler}}\ and\ \bibinfo {author} {\bibfnamefont {C.~H.}\ \bibnamefont
  {Townes}},\ }\href {https://doi.org/10.1103/PhysRev.100.703} {\bibfield
  {journal} {\bibinfo  {journal} {Physical Review}\ }\textbf {\bibinfo {volume}
  {100}},\ \bibinfo {pages} {703} (\bibinfo {year} {1955})}\BibitemShut
  {NoStop}%
\bibitem [{\citenamefont {{Cohen-Tannoudji}}\ and\ \citenamefont
  {Reynaud}(1977)}]{Cohen-Tannoudji1977}%
  \BibitemOpen
  \bibfield  {author} {\bibinfo {author} {\bibfnamefont {C.}~\bibnamefont
  {{Cohen-Tannoudji}}}\ and\ \bibinfo {author} {\bibfnamefont {S.}~\bibnamefont
  {Reynaud}},\ }\href {https://doi.org/10.1088/0022-3700/10/3/005} {\bibfield
  {journal} {\bibinfo  {journal} {Journal of Physics B: Atomic and Molecular
  Physics}\ }\textbf {\bibinfo {volume} {10}},\ \bibinfo {pages} {345}
  (\bibinfo {year} {1977})}\BibitemShut {NoStop}%
\bibitem [{\citenamefont {{Cohen-Tannoudji}}\ \emph {et~al.}(2008)\citenamefont
  {{Cohen-Tannoudji}}, \citenamefont {{Dupont-Roc}},\ and\ \citenamefont
  {Grynberg}}]{Cohen-Tannoudji2008}%
  \BibitemOpen
  \bibinfo {editor} {\bibfnamefont {C.}~\bibnamefont {{Cohen-Tannoudji}}},
  \bibinfo {editor} {\bibfnamefont {J.}~\bibnamefont {{Dupont-Roc}}},\ and\
  \bibinfo {editor} {\bibfnamefont {G.}~\bibnamefont {Grynberg}},\ eds.,\ \href
  {https://doi.org/10.1002/9783527617197} {\emph {\bibinfo {title} {Atom-Photon
  Interactions: Basic Processes and Applications}}},\ A {{Wiley-Interscience}}
  Publication\ (\bibinfo  {publisher} {Wiley},\ \bibinfo {address} {New York,
  NY},\ \bibinfo {year} {2008})\BibitemShut {NoStop}%
\bibitem [{\citenamefont {Beenakker}(1991)}]{Beenakker1991}%
  \BibitemOpen
  \bibfield  {author} {\bibinfo {author} {\bibfnamefont {C.~W.~J.}\
  \bibnamefont {Beenakker}},\ }\href {https://doi.org/10.1103/PhysRevB.44.1646}
  {\bibfield  {journal} {\bibinfo  {journal} {Physical Review B}\ }\textbf
  {\bibinfo {volume} {44}},\ \bibinfo {pages} {1646} (\bibinfo {year}
  {1991})}\BibitemShut {NoStop}%
\bibitem [{\citenamefont {Bush}\ \emph {et~al.}(2021)\citenamefont {Bush},
  \citenamefont {Ochoa},\ and\ \citenamefont {Perron}}]{Bush2021}%
  \BibitemOpen
  \bibfield  {author} {\bibinfo {author} {\bibfnamefont {R.~A.}\ \bibnamefont
  {Bush}}, \bibinfo {author} {\bibfnamefont {E.~D.}\ \bibnamefont {Ochoa}},\
  and\ \bibinfo {author} {\bibfnamefont {J.~K.}\ \bibnamefont {Perron}},\
  }\href {https://doi.org/10.1119/10.0002404} {\bibfield  {journal} {\bibinfo
  {journal} {American Journal of Physics}\ }\textbf {\bibinfo {volume} {89}},\
  \bibinfo {pages} {300} (\bibinfo {year} {2021})}\BibitemShut {NoStop}%
\bibitem [{\citenamefont {Blais}\ \emph {et~al.}(2021)\citenamefont {Blais},
  \citenamefont {Grimsmo}, \citenamefont {Girvin},\ and\ \citenamefont
  {Wallraff}}]{Blais2021}%
  \BibitemOpen
  \bibfield  {author} {\bibinfo {author} {\bibfnamefont {A.}~\bibnamefont
  {Blais}}, \bibinfo {author} {\bibfnamefont {A.~L.}\ \bibnamefont {Grimsmo}},
  \bibinfo {author} {\bibfnamefont {S.~M.}\ \bibnamefont {Girvin}},\ and\
  \bibinfo {author} {\bibfnamefont {A.}~\bibnamefont {Wallraff}},\ }\href
  {https://doi.org/10.1103/RevModPhys.93.025005} {\bibfield  {journal}
  {\bibinfo  {journal} {Reviews of Modern Physics}\ }\textbf {\bibinfo {volume}
  {93}},\ \bibinfo {pages} {025005} (\bibinfo {year} {2021})}\BibitemShut
  {NoStop}%
\bibitem [{\citenamefont {Manzano}(2020)}]{Manzano2020}%
  \BibitemOpen
  \bibfield  {author} {\bibinfo {author} {\bibfnamefont {D.}~\bibnamefont
  {Manzano}},\ }\href {https://doi.org/10.1063/1.5115323} {\bibfield  {journal}
  {\bibinfo  {journal} {AIP Advances}\ }\textbf {\bibinfo {volume} {10}},\
  \bibinfo {pages} {025106} (\bibinfo {year} {2020})}\BibitemShut {NoStop}%
\bibitem [{Note1()}]{Note1}%
  \BibitemOpen
  \bibinfo {note} {Martin et al. designate spin up as their lower energy state,
  so one must also swap up and down indices. They also assume the double
  occupancy state, $b$ never occurs.}\BibitemShut {Stop}%
\bibitem [{\citenamefont {Rams}\ and\ \citenamefont {Zwolak}(2020)}]{Rams2020}%
  \BibitemOpen
  \bibfield  {author} {\bibinfo {author} {\bibfnamefont {M.~M.}\ \bibnamefont
  {Rams}}\ and\ \bibinfo {author} {\bibfnamefont {M.}~\bibnamefont {Zwolak}},\
  }\href {https://doi.org/10.1103/PhysRevLett.124.137701} {\bibfield  {journal}
  {\bibinfo  {journal} {Physical Review Letters}\ }\textbf {\bibinfo {volume}
  {124}},\ \bibinfo {pages} {137701} (\bibinfo {year} {2020})}\BibitemShut
  {NoStop}%
\bibitem [{\citenamefont {Elenewski}\ \emph {et~al.}(2017)\citenamefont
  {Elenewski}, \citenamefont {Gruss},\ and\ \citenamefont
  {Zwolak}}]{Elenewski2017}%
  \BibitemOpen
  \bibfield  {author} {\bibinfo {author} {\bibfnamefont {J.~E.}\ \bibnamefont
  {Elenewski}}, \bibinfo {author} {\bibfnamefont {D.}~\bibnamefont {Gruss}},\
  and\ \bibinfo {author} {\bibfnamefont {M.}~\bibnamefont {Zwolak}},\ }\href
  {https://doi.org/10.1063/1.5000747} {\bibfield  {journal} {\bibinfo
  {journal} {The Journal of Chemical Physics}\ }\textbf {\bibinfo {volume}
  {147}},\ \bibinfo {pages} {151101} (\bibinfo {year} {2017})}\BibitemShut
  {NoStop}%
\bibitem [{\citenamefont {Zedler}\ \emph {et~al.}(2009)\citenamefont {Zedler},
  \citenamefont {Schaller}, \citenamefont {Kiesslich}, \citenamefont {Emary},\
  and\ \citenamefont {Brandes}}]{Zedler2009}%
  \BibitemOpen
  \bibfield  {author} {\bibinfo {author} {\bibfnamefont {P.}~\bibnamefont
  {Zedler}}, \bibinfo {author} {\bibfnamefont {G.}~\bibnamefont {Schaller}},
  \bibinfo {author} {\bibfnamefont {G.}~\bibnamefont {Kiesslich}}, \bibinfo
  {author} {\bibfnamefont {C.}~\bibnamefont {Emary}},\ and\ \bibinfo {author}
  {\bibfnamefont {T.}~\bibnamefont {Brandes}},\ }\href
  {https://doi.org/10.1103/PhysRevB.80.045309} {\bibfield  {journal} {\bibinfo
  {journal} {Physical Review B}\ }\textbf {\bibinfo {volume} {80}},\ \bibinfo
  {pages} {045309} (\bibinfo {year} {2009})}\BibitemShut {NoStop}%
\bibitem [{\citenamefont {{de Vega}}\ and\ \citenamefont
  {Alonso}(2017)}]{deVega2017}%
  \BibitemOpen
  \bibfield  {author} {\bibinfo {author} {\bibfnamefont {I.}~\bibnamefont {{de
  Vega}}}\ and\ \bibinfo {author} {\bibfnamefont {D.}~\bibnamefont {Alonso}},\
  }\bibfield  {journal} {\bibinfo  {journal} {arXiv:1511.06994 [quant-ph]}\
  }\href {https://doi.org/10.1103/RevModPhys.89.15001}
  {10.1103/RevModPhys.89.15001} (\bibinfo {year} {2017}),\ \Eprint
  {https://arxiv.org/abs/1511.06994} {arXiv:1511.06994 [quant-ph]} \BibitemShut
  {NoStop}%
\bibitem [{\citenamefont {Leon}\ \emph {et~al.}(2020)\citenamefont {Leon},
  \citenamefont {Yang}, \citenamefont {Hwang}, \citenamefont {Lemyre},
  \citenamefont {Tanttu}, \citenamefont {Huang}, \citenamefont {Chan},
  \citenamefont {Tan}, \citenamefont {Hudson}, \citenamefont {Itoh},
  \citenamefont {Morello}, \citenamefont {Laucht}, \citenamefont
  {{Pioro-Ladri{\`e}re}}, \citenamefont {Saraiva},\ and\ \citenamefont
  {Dzurak}}]{Leon2020}%
  \BibitemOpen
  \bibfield  {author} {\bibinfo {author} {\bibfnamefont {R.~C.~C.}\
  \bibnamefont {Leon}}, \bibinfo {author} {\bibfnamefont {C.~H.}\ \bibnamefont
  {Yang}}, \bibinfo {author} {\bibfnamefont {J.~C.~C.}\ \bibnamefont {Hwang}},
  \bibinfo {author} {\bibfnamefont {J.~C.}\ \bibnamefont {Lemyre}}, \bibinfo
  {author} {\bibfnamefont {T.}~\bibnamefont {Tanttu}}, \bibinfo {author}
  {\bibfnamefont {W.}~\bibnamefont {Huang}}, \bibinfo {author} {\bibfnamefont
  {K.~W.}\ \bibnamefont {Chan}}, \bibinfo {author} {\bibfnamefont {K.~Y.}\
  \bibnamefont {Tan}}, \bibinfo {author} {\bibfnamefont {F.~E.}\ \bibnamefont
  {Hudson}}, \bibinfo {author} {\bibfnamefont {K.~M.}\ \bibnamefont {Itoh}},
  \bibinfo {author} {\bibfnamefont {A.}~\bibnamefont {Morello}}, \bibinfo
  {author} {\bibfnamefont {A.}~\bibnamefont {Laucht}}, \bibinfo {author}
  {\bibfnamefont {M.}~\bibnamefont {{Pioro-Ladri{\`e}re}}}, \bibinfo {author}
  {\bibfnamefont {A.}~\bibnamefont {Saraiva}},\ and\ \bibinfo {author}
  {\bibfnamefont {A.~S.}\ \bibnamefont {Dzurak}},\ }\href
  {https://doi.org/10.1038/s41467-019-14053-w} {\bibfield  {journal} {\bibinfo
  {journal} {Nature Communications}\ }\textbf {\bibinfo {volume} {11}},\
  \bibinfo {pages} {797} (\bibinfo {year} {2020})}\BibitemShut {NoStop}%
\bibitem [{\citenamefont {Ast}\ \emph {et~al.}(2024)\citenamefont {Ast},
  \citenamefont {Kot}, \citenamefont {Ismail}, \citenamefont
  {{de-la-Pe{\~n}a}}, \citenamefont {{Fern{\'a}ndez-Dom{\'i}nguez}},\ and\
  \citenamefont {Cuevas}}]{Ast2024}%
  \BibitemOpen
  \bibfield  {author} {\bibinfo {author} {\bibfnamefont {C.~R.}\ \bibnamefont
  {Ast}}, \bibinfo {author} {\bibfnamefont {P.}~\bibnamefont {Kot}}, \bibinfo
  {author} {\bibfnamefont {M.}~\bibnamefont {Ismail}}, \bibinfo {author}
  {\bibfnamefont {S.}~\bibnamefont {{de-la-Pe{\~n}a}}}, \bibinfo {author}
  {\bibfnamefont {A.~I.}\ \bibnamefont {{Fern{\'a}ndez-Dom{\'i}nguez}}},\ and\
  \bibinfo {author} {\bibfnamefont {J.~C.}\ \bibnamefont {Cuevas}},\
  }\href@noop {} {\bibinfo {title} {Theory of {{Electron Spin Resonance}} in
  {{Scanning Tunneling Microscopy}}}} (\bibinfo {year} {2024}),\ \Eprint
  {https://arxiv.org/abs/2403.20247} {arXiv:2403.20247 [cond-mat]} \BibitemShut
  {NoStop}%
\bibitem [{\citenamefont {Paul}\ \emph {et~al.}(2016)\citenamefont {Paul},
  \citenamefont {Baumann}, \citenamefont {Lutz},\ and\ \citenamefont
  {Heinrich}}]{Paul2016}%
  \BibitemOpen
  \bibfield  {author} {\bibinfo {author} {\bibfnamefont {W.}~\bibnamefont
  {Paul}}, \bibinfo {author} {\bibfnamefont {S.}~\bibnamefont {Baumann}},
  \bibinfo {author} {\bibfnamefont {C.~P.}\ \bibnamefont {Lutz}},\ and\
  \bibinfo {author} {\bibfnamefont {A.~J.}\ \bibnamefont {Heinrich}},\ }\href
  {https://doi.org/10.1063/1.4955446} {\bibfield  {journal} {\bibinfo
  {journal} {Review of Scientific Instruments}\ }\textbf {\bibinfo {volume}
  {87}},\ \bibinfo {pages} {074703} (\bibinfo {year} {2016})}\BibitemShut
  {NoStop}%
\end{thebibliography}%

\end{document}